\documentclass[authoryear]{elsarticle}
\usepackage{amsmath}
\usepackage{amsthm}
\usepackage{amssymb}
\usepackage{mathtools}
\usepackage{algorithm}
\usepackage{algorithmic}
\usepackage{url}

\RequirePackage[colorlinks,citecolor=blue,urlcolor=blue]{hyperref}

\def\bs{\boldsymbol}
\def\ov{\overline}

\def\v{v_\alpha}
\def\A{{\cal A}}
\def\E{{\mathbb E}}                         % Expectation
\def\V{{\mathbb V}}
\def\Q{\mathcal{Q}}
\def\S{\mathcal{S}}
\def\AIS_smc{density tempered sequential Monte Carlo}
\def\dt_SMC{density tempered SMC}
\def\Dt_SMC{Density tempered SMC}
\def\dtsmc{density tempered sequential Monte Carlo}
\renewcommand{\eqref}[1]{Eq.~(\ref{#1})}
\newcommand{\secref}[1]{Section~\ref{#1}}

\newcommand{\algref}[1]{Algorithm~\ref{#1}}

\begin{document}

\begin{frontmatter}

\title{New Estimation Approaches for the Hierarchical Linear Ballistic Accumulator Model}

\author[uow,acems]{D. Gunawan}
\ead{dgunawan@uow.edu.au}
\author[uon]{G. E. Hawkins}
\ead{guy.hawkins@newcastle.edu.au}
\author[usyd,acems]{M.-N. Tran}
\ead{minh-ngoc.tran@sydney.edu.au}
\author[unsw,acems]{R. Kohn}
\ead{r.kohn@unsw.edu.au}
\author[uon]{S. D. Brown}
\ead{scott.brown@newcastle.edu.au}

\address[unsw]{School of Economics, UNSW Business School}
\address[uon]{School of Psychology, University of Newcastle}
\address[usyd]{Discipline of Business Analytics, University of Sydney Business School}
\address[acems]{ARC Centre of Excellence for Mathematical and Statistical Frontiers (ACEMS)}
\address[uow]{School of Mathematics and Applied Statistics, University of Wollongong}

\sloppy
\begin{abstract}
The Linear Ballistic Accumulator \cite[LBA:][]{Brown2008} model is used as a measurement tool to answer questions about applied psychology. 
The analyses based on this model depend upon the model selected and its estimated parameters. Modern approaches use hierarchical Bayesian 
models and Markov chain Monte-Carlo (MCMC) methods to estimate the posterior distribution of the parameters. 
Although there are several approaches available for model selection, they are all based on the posterior samples produced via MCMC, 
which means that the model selection inference inherits the properties of the MCMC sampler. 
To improve on current approaches to LBA inference we propose two methods that  
are based on recent advances in particle MCMC methodology; they are qualitatively different from existing approaches as well as 
from each other. The first approach is particle Metropolis-within-Gibbs; the second approach is density tempered sequential Monte Carlo.
%\dtsmc\footnote{MNT: are you OK with this or do you want to go back to 
% AISIL-RE or Annealed Importance Sampling with intractable 
%likelihood and Random Effects. It seems to me that density tempered sequential Monte Carlo, or density tempered SMC  seems shorter 
%and more descriptive, but it's just a question of taste }.
Both new approaches provide very efficient sampling and can be applied to estimate the marginal likelihood, which provides Bayes factors for model selection. 
The first approach is usually faster. The second approach 
provides a direct estimate of the marginal likelihood, uses the first approach in its Markov move step and is very efficient to parallelize on high performance computers.  The new methods are illustrated by applying them to simulated and real data, and through pseudo code.  
The code implementing the methods is freely available. 
\end{abstract}


\end{frontmatter}

\section{Introduction\label{Sec: introduction}}
The Linear Ballistic Accumulator (LBA; \citealp{Brown2008}) provides a tractable model of decision making
which is simpler than some other models of choice response time because it eliminates complexities such as
competition between alternatives \citep{Brown2005,Ratcliff1978,RatcliffRouder1998}, and passive decay of
evidence \citep{RatcliffSmith2004,Usher2001}. The model's simplicity allows analytic solutions for 
choices between any number of alternatives. Like other evidence accumulation models, the LBA model is 
used to address theoretical and applied questions about human cognition, both in the general population and in
clinical groups \citep[for reviews, see e.g., ][]{donkin2018response,ratcliff2016diffusion}.

When used  as a psychometric tool, key inferences are drawn in two ways: either from parameter estimates or by comparing different versions of the LBA model estimated from the same data. These inferences
rely on accurate parameter estimates and valid model selection procedures, which can be difficult problems.
Most modern applications of the model use hierarchical structures estimated in a Bayesian framework, with
posterior distributions over the parameters estimated using Markov chain Monte Carlo (MCMC). In almost all
Bayesian applications of the LBA model, the MCMC uses Metropolis steps, with proposals generated by differential
evolution \citep[DE-MCMC:][]{Turner2013}.\footnote{Although see \citet{Annis2017} for sampling 
with a no-U-turn sampler, in STAN. That approach may be suitable for some smaller problems.}%\footnote{Do Annis et al. say that Stan works poorly or do we need to say that here?
%DG response: Of course Annis did not say that STAN did not work in their paper because it is their contribution.
%RK: I REMEMBER WE HAD SOME EVIDENCE THAT LBA IN STAN DID NOT WORK. DO YOU REMEMBER WHAT IT WAS ?
%}
%DE-MCMC alleviates some of the challenging aspects of parameter estimation
%in the LBA, caused by the substantial structural correlations between the model parameters. 
After sampling from
the posterior distribution, inferences about model selection are almost always carried out by estimating a
marginal likelihood, or some quantity that behaves approximately like the marginal likelihood, from the MCMC
samples. Commonly used model selection metrics include the deviance information criterion
\citep[DIC:][]{spiegelhalter2014deviance} and the Watanabe (or ``widely applicable'') information criterion
\citep[WAIC:][]{watanabe2010asymptotic}. Model selection using Bayes factors or estimated Bayes factors
requires estimating the marginal likelihood of each competing model. \citet{evans2018bayes} estimate the
marginal likelihood by generating a large number of samples from the prior. This method requires specialised computing hardware (a general purpose graphical processing unit) to be computationally feasible even in smaller, non-hierarchical applications. However, obtaining reliable and stable estimation of the marginal likelihood of the hierarchical LBA model remains challenging. 

\citet{Gronau:2019} and \citet{Evans:2019} propose estimating the marginal likelihood by bridge sampling and thermodynamic integration respectively, from the MCMC samples. These two methods are promising even for the hierarchical LBA model, but depend on the quality of the MCMC samples. Both \citet{Gronau:2019} and \citet{Evans:2019} use the DE-MCMC algorithm to sample from the posterior distribution of the parameters, which 
often still suffers from the usual problems associated with random walk samplers for the hierarchical LBA model with a 
large number of parameters, including a high autocorrelation between samples, and slow or uncertain convergence. When the DE-MCMC samples provide an imperfect
representation of the posterior, the subsequent model selection methods using bridge sampling or thermodynamic integration will give incorrect estimates. Both
bridge sampling and thermodynamic integration would benefit from a more reliable and efficient sampling algorithm to obtain reliable estimates of the marginal likelihood for model comparison.
Furthermore, bridge sampling uses an iterative algorithm to estimate the marginal likelihood; \citet{Gronau:2019}
noted that in rare cases, a very good starting value for the marginal
likelihood is crucial for the algorithm to converge. Thermodynamic integration also requires the MCMC algorithm to sample efficiently 
at different temperatures, which can be difficult to tune. Section \ref{MarginalLikelihoodEstimation} further discusses the thermodynamic integration method. It is clear that more efficient and robust sampling methods could be beneficial in many ways.

%\section{Introduction}

%It is out of the scope of our article to carefully study the various methods for estimating the marginal likelihood. 
%we propose a modified parameterisation of the hierarchical LBA model which allows the individual level parameters
%to be correlated with each other \textit{a priori}.
Our paper makes three substantive contributions. First, we allow the individual level parameters to be correlated in the prior by reparameterising them. An essential component to this parameterisation is to first transform the individual level parameters so they can take all values on the real line which makes it straightforward to specify their joint distribution as an unconstrained multivariate normal with full covariance matrix structure. Previous approaches followed
\citet{Turner2013} and assumed that the prior joint distribution of these parameters are uncorrelated truncated (positive only) univariate normal distributions. 
The new assumptions and parameterisation have the twin advantages of more accurately reflect prior knowledge, and increasing sampling efficiency. Furthermore, 
Section~\ref{sec:Simulation-Study-and real applications} shows that using a hierarchical LBA model with uncorrelated univariate normal distributions 
for the transformed random effects can result in overconfidence in estimation precision and underestimation of the magnitude of the individual differences. That section also explains why it is difficult or maybe impossible to efficiently apply the existing DE-MCMC sampler to the new parameterisation 
of the hierarchical LBA model. 

The second and third contributions propose two new methods for estimating the improved LBA model 
that are more efficient than the DE-MCMC sampler used for the estimation of the LBA since \citet{Turner2013}.
The first method is based on the particle Metropolis within Gibbs (PMwG) approach of \citet{Gunawan2017}.
It defines a target posterior density on an augmented space that includes the standard model parameters as well as
multiple copies of the individual random effects (``particles'') and whose marginal density is the joint
posterior density of the parameters and random effects. Section~\ref{sec:Simulation-Study-and real
applications} shows that for estimating the LBA model, the PMwG sampler is an important alternative to the
DE-MCMC sampler because it converges much more reliably and is much more statistically efficient.

The second estimation method is based on a version of sequential Monte Carlo (SMC; \citealp{DelMoral:2006}) that 
is an alternative to all the MCMC approaches mentioned above, including DE-MCMC and PMwG. 
Our approach builds on the work by \citet{Neal:2001}, \citet{DelMoral:2006} and \citet{Duan2015} by first 
drawing samples from an easily-generated distribution, such as the multivariate normal prior, and then moving those samples sequentially 
towards the posterior distribution. We call our algorithm \dtsmc{} (DT-SMC); it uses 
three main steps to transition from one 
intermediate density to the next: a reweighting step that 
moves the particles from one tempered density to the next; a resampling step
that eliminates particles with low weights; and the Markov step that applies several iterates of the new PMwG sampler 
to help ensure that the particles represent the tempered target density adequately.

Section~\ref{sec:Simulation-Study-and real applications} shows that PMwG is much faster than 
\dt_SMC{}; it is also  easily implemented with very modest computational resources, e.g. personal computers. However,
\dt_SMC{} is easier to parallelize than PMwG, so that it is likely to be faster when powerful computing resources are available. The \dt_SMC{} explores the parameter space more
efficiently when the target posterior distribution is multimodal; such distributions are usually due to multimodal priors with small sample sizes. Both
PMwG and \dt_SMC{} can be used to estimate the marginal likelihood; the marginal likelihood can be obtained from the \dt_SMC{} algorithm with negligible extra cost, while the output from the PMwG sampler can be used by 
both thermodynamic integration and bridge sampling. We believe that 
it is important to present both the PMwG and \dt_SMC{} algorithms to give users the ability to estimate LBA models 
using both modest and considerable computational resources; in addition, future improvements in 
both bridge sampling and thermodynamic integration may make it attractive to estimate the marginal likelihood by first
running PMwG.  However, it is outside the scope of the article to carefully 
compare the performance of all current approaches for estimating the marginal likelihood. 
%\footnote{all: I tried above to say that: 
%\begin{itemize}
%    \item we have two methods to estimate the posterior; nothing about marginal likelihood yet. 
%    \item 
%    gave their strengths and weaknesses
%    \item said both methods can be used to estimate the marginal likelihood. 
%    \item 
%   said that both methods should be presented in the paper and gave reasons. I think this is the weakest part of the argument. 
%\end{itemize}
%any suggestions on improvement. 
%}

The rest of the paper is organised as follows. Section \ref{sec:The-Linear-Ballistic accumulator}
describes the Linear Ballistic Accumulator model; Section \ref{sec:Bayesian-Estimation}
presents the Bayesian estimation methodologies; 
Section \ref{sec:Simulation-Study-and real applications} discusses the estimation results
where the two new methods are applied to simulated and real data; and Section \ref{sec:Conclusions} concludes.
The paper has several appendices which contain some further implementation details and technical results.
An online supplement containing some further empirical and technical results and code applying the two 
estimation methods to an example dataset is available at \url{osf.io/5b4w3}.

\section{The Linear Ballistic Accumulator (LBA) model}\label{sec:The-Linear-Ballistic accumulator}

%% Do we need this material?
% For a single participant, the $i^{th}$ observation in a choice experiment will contain two pieces of information: the response choice, which we denote $RE_{i}\in\left\{ 1,...,C\right\} $, where $C$ is the number of response alternatives; and the response time, which we denote $RT_{i}\in\left(0,\infty\right)$.

To more precisely discuss the algorithms with the updated model specification, we use a slightly different notation for the LBA than previous literature has used. Usually, the LBA model represents a choice between $C$ alternatives $\left(C=2,3,...\right)$ using $C$ different evidence accumulators, one for each response choice. 
Each accumulator begins with an independent amount of starting evidence $k_{c}$ which is sampled independently for each accumulator from a continuous uniform distribution $k_{c}\sim U\left(0,A\right)$. The evidence for accumulator $c$
increases at a drift rate $d_{c}$ which is sampled independently for each accumulator from
a normal distribution with mean $v_{c}$ and standard deviation $s$,
so $d_{c}\sim N\left(v_{c},s\right)$,
although other non-normal distributions are possible \citep{terry2015generalising}. To satisfy the scaling conditions of the model, it
is common to set the variance of the sampled drift rates to one, i.e., $s=1$;
however, see also \citet{DonkinBrownHeathcote2009}. Each accumulator gathers evidence
until one accumulator reaches a response threshold $b$. The LBA model
assumes that the observed response time $RT$ is the sum of the decision time, plus
some extra time $\tau$ for the non-decision process such as stimulus encoding and motor execution. 
For simplicity, $\tau$ is usually assumed to be constant
across trials. Thus, the final observed %$RT$ is given by
\[
RT=\underset{c}{\min}\left\{\frac{b-k_{c}}{d_{c}}+\tau\right\}.
\]
Let $T_c=(b-k_c)/d_c+\tau$ be the time for accumulator $c$ to reach the threshold $b$. 
\citet{Brown2008} derive the cumulative distribution  function of $T_c$ as
\begin{equation*}
\begin{aligned}
F_c(t)&=&1+\frac{b-A-(t-\tau)v^c}{A}\Phi\left(\varpi_1-\varpi_2\right)-\frac{b-(t-\tau)v^c}{A}\Phi\left(\varpi_1\right)\\
&&+\frac{1}{\varpi_2}\phi\left(\varpi_1-\varpi_2\right)-\frac{1}{\varpi_2}\phi\left(\varpi_1\right),
\end{aligned}
\end{equation*}
and its density as 
\begin{equation*}\label{LBAdens}
f_c(t)=\frac{1}{A}\Big[-v^c\Phi\left(\varpi_1-\varpi_2\right)+s\phi\left(\varpi_1-\varpi_2\right)+v^c\Phi\left(\varpi_1\right)-s\phi\left(\varpi_1\right)\Big]; 
\end{equation*}
above, $\phi$ and $\Phi$ are the density and cumulative distribution functions of the standard normal distribution, respectively, and 
$$\varpi_1=\frac{b-(t-\tau)v^c}{(t-\tau)s}\quad {\rm and}\quad \varpi_2=\frac{A}{(t-\tau)s}.$$

%If the observed response choice is $RE = c $ and $RT  \in (t, t+ \d t ) $, 
%where $\delta t $ is a very small time interval, then
%\[\Pr(RE=c,RT\in (t,t+\d t))=p(T_c\in (t,t+\d t),T_k>t, k\not= c)\approx f_c(t)\d t\times \prod_{k\not= c}(1-F_k(t)).\]
It follows that the joint density of the response choice $RE$
and response time $RT$ at $(RE,RT)=(c,t)$, given the values $b,A,v,s,\tau,$ is defined as
\begin{align}\label{eq: joint density RE RC}
\text{LBA}(c,t|b,A,v,s,\tau):=f_c(t)\times \prod_{k\not= c}(1-F_k(t)). 
\end{align}
\ref{DetailjointdensityRTandRE} gives technical details of the joint density of the response choice $RE$
and response time $RT$.  The supplement at \url{osf.io/5b4w3} shows that the LBA density  \eqref{eq: joint density RE RC}
is bounded.

If a subject makes $N$ independent decisions, with choices $RE_i, i=1\dots N,$ 
and corresponding response times $RT_i$, the density of $\bs {RE}:=(RE_1,\dots, RE_N)$ and 
$ \bs {RT}:=(RT_1,\dots, RT_N)$ is
\[
p\left(\bs {RE,RT} |b,A,v,s,\tau\right)=\prod_{i=1}^{N}\textrm{LBA}\left(RE_{i},RT_{i}|b,A,v,s,\tau\right).
\]
In applications of the LBA model, it is possible that the parameters ($b$, $A$, etc.) vary over different conditions of the
experiment, and sometimes also across the different accumulators.

%\subsection{Hierarchical Bayesian implementation of the LBA model}\label{subsec:Hierarchical-Bayesian-Model}

The setup for the hierarchical LBA model is motivated by data collected from the decisions of 19
young subjects and first presented by \citet{Forstmann2008}. 
The participants were asked to decide, repeatedly, whether a cloud of semi-randomly moving dots appeared to move to the left or to the right. Before each decision trial, subjects were instructed about what quality of their decision-making they should emphasise. For some trials, they were asked to respond as accurately as possible, for other trials they were asked to respond at their own pace, and for other trials they were asked to respond as quickly as possible. We label these conditions, in order: ``accuracy emphasis'' (condition 1); ``neutral emphasis'' (condition 2); and ``speed emphasis'' (condition 3). The different conditions were randomly mixed from trial to trial, with the subjects cued by a word which appeared on screen before each decision stimulus. Let $Z$ be the number of conditions in the experiment, with $Z=3$ here. Each subject made $N_{z}=280$ decisions for $z=1,...,Z$ ($N=840$ trials in total). See \citet{Forstmann2008} for more details on the procedure and the data, including the associated neuroimaging measurements, which are not considered here.

To model the differences between the three conditions in the experiment, we follow \citet{Forstmann2008} and allow
different threshold parameters $b^{\left(1\right)}$, $b^{\left(2\right)}$ and $b^{\left(3\right)}$
for the accuracy, neutral and speed conditions, respectively. $RE_{i,j}$ and $RT_{i,j}$ denote the $i$th response from the $j$th
subject. Following \citeauthor{Forstmann2008}, we also collapse data across right-moving and left-moving
stimuli, and so we index means of the drift rate distributions as $ v^{\left(1\right)}$ and $v^{\left(2\right)}$ for the accumulator
corresponding to incorrect and correct response choice, respectively. We assume that the standard deviation of the drift rate distribution is always $s=1$. Together, these assumptions imply
that each subject $j$ has the vector of individual-level parameters or random effects
\[\left(b_{j}^{\left(1\right)},b_{j}^{\left(2\right)},b_{j}^{\left(3\right)},A_{j},\tau_{j},v_{j}^{\left(1\right)},v_{j}^{\left(2\right)}\right),\;\; j=1,...,S.
\]

With the usual assumptions of independence, the conditional density of all the observations is
% \footnote{\begin{itemize}
%     \item 
% David, please check the next equation for correct $b_j^{z}$ and $v_j^{\cdot}$ as I am not checking for mathematical accuracy. \item 
% Also, if we have $N$ trials for each individual, then i don't see where the extra $\prod_{z=1}^Z$ comes from. 
% \end{itemize}
% }
\begin{equation}
p\left(\bs {RT,RE}|\boldsymbol{b},\boldsymbol{A},\boldsymbol{\tau},\boldsymbol{v}\right)=\prod_{j=1}^{S}\prod_{z=1}^{Z}\prod_{i=1}^{N_{z}}\textrm{LBA}\left(RE_{i,j,z},RT_{i,j,z}|b_{j}^{\left(z\right)},A_{j},v_{j}^{\left(1\right)},v_{j}^{\left(2\right)},\tau_{j}\right).\label{eq:conditional likelihood Hierarchical LBA}
\end{equation}
 %Each of the individual random effects is restricted to be positive (it is theoretically possible, but psychologically implausible, for the means of the drift rate distributions, $v_j$, to be negative). Respecting this,
 Each of the individual fixed effects is restricted to be positive (it is theoretically possible, but psychologically implausible, for the means of the drift rate distributions, $v_j$, to be negative). Respecting this,
\citet{Turner2013} specified uncorrelated truncated normal distributions
for each of these individual level parameters, and this has become standard in hierarchical applications of the LBA since then.
\citet{Turner2013} and \cite{evans2018modeling} also found that the posterior distributions of the
individual random effects are highly correlated. Despite this, it has been standard practice to model the random effects as being \textit{a priori} independent.
% \footnote{what we had before is 
% ``Despite this, standard practice has been to specify
% independent distributions for each of them" which does not make that much sense. }. 

To improve both the computational efficiency of the algorithms and inference accuracy,
we use instead a hierarchical model based on a multivariate normal
distribution of log-transformed random effects, with an explicitly-estimated variance;  
implying that the group distribution on the non-transformed scale is multivariate log-normal. 
For each subject $j=1, \dots, S$, we define the vector of random effects
\begin{equation}\label{eq: subject RE}
\bs{\alpha}_{j} =(\alpha_{1j}, \cdots, \alpha_{7j}) :=\log \Big(b_{j}^{\left(1\right)},b_{j}^{\left(2\right)},b_{j}^{\left(3\right)},A_{j},
v_{j}^{\left(1\right)},v_{j}^{\left(2\right)},\tau_{j}\Big)%=:\Big(\alpha_{b_{j}^{\left(1\right)}},\alpha_{b_{j}^{\left(2\right)}},\alpha_{b_{j}^{\left(3\right)}},\alpha_{A_{j}},
\end{equation}
Throughout, we use short-hand such as $\alpha_{1,j}$ to refer to the first element of the log-transformed parameter vector for participant $j$, and $\bs \alpha_{j}$ to refer to the corresponding random effects vector for participant $j$. 
The dimension of $\bs \alpha_j$ is $D_\alpha = 7$ here. 
Not all users may want to constrain the mean drift rates to be strictly positive, and so of course those may be left out of the log transformation in some cases.

%\begin{equation}\label{eq: subject RE}
%\bs{\alpha}_{j} =\Big(\alpha_{b_{j}^{\left(1\right)}},\alpha_{b_{j}^{\left(2\right)}},\alpha_{b_{j}^{\left(3\right)}},\alpha_{A_{j}},
%\alpha_{v_{j}^{\left(1\right)}},\alpha_{v_{j}^{\left(2\right)}},\alpha_{\tau_{j}}\Big),
%\end{equation}
%where $\alpha_{b_{j}^{\left(z\right)}} =\log\big(b_{j}^{\left(z\right)}\big)$,
%$\alpha_{A_{j}}=\log\left(A_{j}\right)$, $\alpha_{v_{j}^{\left(c\right)}}
%=\log\big(v_{j}^{\left(c\right)}\big)$ and $\alpha_{\tau_{j}}=\log\left(\tau_{j}\right)$.
%Let $D_\alpha = 7$ be the dimension of $\bs \alpha_j$. Not all users may want to constrain the mean drift rates to be strictly positive, and so of course those may be left out of the log transformation in some cases. 

To account for the dependence between the random effects, the prior distribution of the vector $\bs \alpha_j$ is modelled as
%\footnote{there is no need to have a subscript on $\mu$ and $\Sigma$}
\begin{equation} \label{eq: prior on alpha_j}
\bs{\alpha}_{j}|\bs \mu,\bs \Sigma  \sim N\left(\bs \mu,\bs \Sigma\right) .
\end{equation}

%\begin{equation} \label{eq: prior on alpha_j}
%\bs{\alpha}_{j}|\bs \mu_\alpha,\bs \Sigma_\alpha  \sim N\left(\bs \mu_\alpha,\bs \Sigma_\alpha\right) .
%\end{equation}

There are a number of priors in the literature available for the parameters $\bs \mu$ and $\bs \Sigma$. We take the normal $N\left(0,I_{D_\alpha}\right)$ prior for $\bs \mu$,
and the marginally non-informative prior of \citet{Huang2013} for
$\bs \Sigma$,
\begin{equation}
\begin{aligned}\label{eq:wandprior}
\bs \Sigma|a_{1},...,a_{D_\alpha} & \sim  IW\left(\v+D_\alpha-1,2\v\textrm{diag}\left(1/a_{1},...,1/a_{D_\alpha}\right)\right),\\
a_{d}& \sim  IG\left(\frac{1}{2},\frac{1}{\A_{d}^{2}}\right),d=1,...,D_\alpha,
\end{aligned}
\end{equation}
where $\v$, $\A_{1}$,...,$\A_{D_\alpha}$
are positive scalars and $\textrm{diag}\left(1/a_{1},...,1/a_{D_\alpha}\right)$
is a diagonal matrix with diagonal elements $1/a_{1},...,1/a_{D_\alpha}$.
%, where $D_\alpha$ is the dimension of $\bs{\alpha}_{j}$.
The notation $IW(a,A)$ means an inverse Wishart  distribution with degrees of freedom $a$
and scale matrix $A$ and the notation
$IG(a,b)$ means an  inverse Gamma distribution with scale parameter $a$ and shape parameter $b$.
We choose this prior for $\bs \Sigma$ because it leads to psychologically plausible marginal prior distributions for the elements of the covariance matrix. \citeauthor{Huang2013} show that \eqref{eq:wandprior} induces $\textrm{half-t}\left(\v,\A_{d}\right)$ distributions for each standard deviation term in $\bs \Sigma$ and setting $\v=2$ leads to marginally
uniform distributions for all the correlation terms in $\bs \Sigma$.
In our application, we set $\v=2$ and $\A_d=1$ for all $d=1,\dots, D_\alpha$. These prior densities cover most possible values in practice, and are relatively non-informative.
The specification we have used implies that the distribution for the random effects vector $\exp(\bs{\alpha}_{j}) |\bs\mu,\bs \Sigma$
is a multivariate log-normal distribution with mean and covariance matrix given by
\begin{equation} \label{eq: log normal distn}
  \bs{\tilde\mu}  = \E\big (\exp(\bs \alpha)|\bs \mu, \bs \Sigma \big )
\quad \text{and} \quad \bs{\tilde\Sigma} = \V \big (\exp (\bs \alpha)|\bs \mu, \bs \Sigma  \big )
\end{equation}
so that $\tilde\mu_{i} =\exp\big(\mu_{i}+\frac{1}{2}\Sigma_{ii}\big)$,
$\tilde\Sigma_{ik} =\exp\big(\mu_{i}+\mu_{k}+0.5\left(\Sigma_{ ii}+\Sigma_{kk}\right)\big)\big(\exp(\Sigma_{ik})-1\big)$.
%The specification we have used implies that the distribution for the random effects vector $\exp(\bs{\alpha}_{j}) |\bs\mu_\alpha,\bs \Sigma_\alpha$
%is a multivariate log-normal distribution with mean and covariance matrix given by
%\begin{equation} \label{eq: log normal distn}
%\bs \mu_{LN,\alpha}  = \E\big (\exp(\bs \alpha)|\bs \mu_\alpha, \bs \Sigma_\alpha \big )
%\quad \text{and} \quad \bs \Sigma_{LN, \alpha} = \V \big (\exp (\bs \alpha)|\bs \mu_\alpha, \bs \Sigma_\alpha  \big )
%\end{equation}
%so that $\mu_{LN,\alpha,i}:= (\bs \mu_{LN,\alpha})_i =\exp\Big(\mu_{\alpha, i}+\frac{1}{2}\Sigma_{\alpha, ii}\big)$,
%$\Sigma_{LN,\alpha,ik} := (\bs\Sigma_{LN, \alpha} )_{ik}  =\exp\big(\mu_{\alpha, i}+\mu_{\alpha, k}+0.5\left(\Sigma_{\alpha, ii}+\Sigma_{\alpha,kk}\right)\big)\big(\exp(\Sigma_{\alpha, ik})-1\big)$.

%The hierarchical prior given by \eqref{eq: subject RE} - \eqref{eq:wandprior} has some important modeling  advantages over the previous settings in the LBA literature because it allows the random effects to be a priori correlated, with  the  parameters of the prior estimated from the data. The second improvement is that we use the marginally non-informative prior of \citet{Huang2013} for the variance-covariance matrix as opposed to usual inverse Wishart prior. The extra information assumed in the inverse Wishart prior may not always be well justified.

\section{Bayesian Estimation}\label{sec:Bayesian-Estimation}
This section develops efficient Bayesian inference for the
hierarchical LBA model described in \secref{sec:The-Linear-Ballistic accumulator}. We use the particle MCMC approach of
\cite{Gunawan2017} and also develop a density tempered SMC approach which relies on particle MCMC.

%\subsection{Preliminaries\label{SS; preliminaries}}
Let $\boldsymbol{\theta}\in\boldsymbol{\Theta}\subset \mathbb{R}^{d_{\theta}}$ be the vector of unknown group-level parameters, $d_{\theta}$ the dimension of the parameters, and $p(\bs \theta)$ be the prior for $\bs \theta$, where $\mathbb{R}^m$ is $m$ dimensional Euclidean
space for a positive integer $m$. 
Let $\bs y_j$ be the vector of observations for the $j$th subject,
and define $\boldsymbol{y}=\boldsymbol{y}_{1:S}=\left(\boldsymbol{y}_{1},...,\boldsymbol{y}_{S}\right)$ as the vector of observations for all $S$ subjects.
Let $\bs \alpha_j \in \bs\chi \subset \mathbb{R}^{D_{\bs \alpha}}$ be the vector of individual-level parameters (random effects) 
for subject $j$,
and $p(\bs \alpha_j|\bs \theta)$ its density under the group-level distribution. $\Theta$ and $\bs\chi$ are themselves Euclidean spaces in all cases we consider. 
Now define 
$\boldsymbol{\alpha}=\boldsymbol{\alpha}_{1:S}=\left(\boldsymbol{\alpha}_{1},...,\boldsymbol{\alpha}_{S}\right)$ 
as the vector of all individual-level parameters, whose dimension is $D_{\bs \alpha}$.

We assume that the $\bs \alpha_j$ 
are independent a priori given $\bs \theta$ and that the ${\bs y}_j$ are independent 
given $\bs \theta$ and $\bs \alpha$, i.e.,
%\begin{align*}
\begin{equation}
p(\bs \alpha_{1:S}|\bs \theta)  = \prod_{j=1}^S p(\bs \alpha_j|\bs \theta) \quad \text{and} \quad
p(\bs y| \bs \theta , \bs \alpha_{1:S}) = \prod_{j=1}^S p(\bs y_{j}| \bs \theta , \bs \alpha_{j})
\end{equation}

Our goal is to obtain samples from the posterior density 
\begin{equation}
p\left(\boldsymbol{\theta},\boldsymbol{\alpha}_{1:S}\right|\boldsymbol{y})\coloneqq p\left(\boldsymbol{y}_{1:S}|\boldsymbol{\theta},\boldsymbol{\alpha}_{1:S}
\right)p\left(\boldsymbol{\alpha}_{1:S}|\boldsymbol{\theta}\right)p\left(\boldsymbol{\theta}\right)/p\left(\boldsymbol{y}_{1:S}\right),\label{eq:posterior}
\end{equation}
where
\begin{equation}
p\left(\boldsymbol{y}\right)=\iint p\left(\boldsymbol{y}_{1:S}|\boldsymbol{\theta},\boldsymbol{\alpha}_{1:S}\right)p\left(\boldsymbol{\alpha}_{1:S}|
\boldsymbol{\theta}\right)p\left(\boldsymbol{\theta}\right)d\boldsymbol{\theta}d\boldsymbol{\alpha}_{1:S}\label{eq:marginal likelihood}
\end{equation}
is the marginal likelihood used in Bayesian inference
to choose between competing models; see, e.g.,  \citet{Kass:1995,Chib2001}.

%\secref{subsec:Hierarchical-Bayesian-Model} describes how to compute the densities

\eqref{eq:conditional likelihood Hierarchical LBA}, \eqref{eq: prior on alpha_j} and \eqref{eq:wandprior} in Section \ref{sec:The-Linear-Ballistic accumulator} describe the densities $p\left(\boldsymbol{y}_{1:S}|\boldsymbol{\theta},\boldsymbol{\alpha}_{1:S}\right)$,
$p\left(\boldsymbol{\alpha}_{1:S}|\boldsymbol{\theta}\right)$ and
$p\left(\boldsymbol{\theta}\right)$, respectively.
We are usually also
interested in estimating posterior distributions of functions
$\varphi\left(\boldsymbol{\theta},\boldsymbol{\alpha}_{1:S}\right)$
and their posterior expectations, 
i.e.,
\begin{equation}
\E\left(\varphi\right)=\iint\varphi\left(\boldsymbol{\theta},\boldsymbol{\alpha}_{1:S}\right)
p\left(\boldsymbol{\theta},\boldsymbol{\alpha}_{1:S}\right|\boldsymbol{y})d\boldsymbol{\theta}d\boldsymbol{\alpha}_{1:S},\label{eq:expectation functions}
\end{equation}
as well as estimating the marginal likelihood in \eqref{eq:marginal likelihood}.
%which is used in model selection.

\subsection{Particle Markov chain Monte Carlo (PMCMC)}\label{subsec:Particle-Markov-chain Monte Carlo}
The particle Metropolis with Gibbs (PMwG) sampler of  \citet{Gunawan2017} is used
for MCMC sampling. This sampler defines a target distribution on an augmented space 
that includes the model parameters
and multiple copies of the individual random effects (``particles''). 

Let $\left\{ m_{j}\left(\boldsymbol{\alpha}_{j}|\boldsymbol{\theta},
\boldsymbol{y}_{j}\right);j=1,...,S\right\} $
be a family of proposal densities that is used to approximate the conditional
densities $\left\{ p\left(\boldsymbol{\alpha}_{j}|\boldsymbol{\theta},\boldsymbol{y}_{j}\right);j=1,...,S\right\} $. \ref{AssumptionProp} gives the technical assumptions required for these proposal densities.

Let $\boldsymbol{\alpha}_{j}^{r}$ be the $r$th sample from the proposal density
$m_{j}\left(\boldsymbol{\alpha}_{j}|\boldsymbol{\theta},\boldsymbol{y}_{j}\right)$ for subject $j$. 
Define $\boldsymbol{\alpha}_{1:S}^{1:R}\coloneqq\left\{ \boldsymbol{\alpha}_{1}^{1:R},...,\boldsymbol{\alpha}_{S}^{1:R}\right\} $
and $\boldsymbol{\alpha}_{j}^{1:R}\coloneqq\big\{ \boldsymbol{\alpha}_{j}^{1},...,\boldsymbol{\alpha}_{j}^{R}\big\}$.
Then the joint density of the particles $\boldsymbol{\alpha}_{1:S}^{1:R}$ based on these proposals, and conditional on $\boldsymbol{\theta}$ and $\boldsymbol{y}$,  is
\begin{equation}
\psi\left(\boldsymbol{\alpha}_{1:S}^{1:R}|{\boldsymbol{\theta}},\boldsymbol{y}\right)=\prod_{r=1}^{R}\prod_{j=1}^{S}m_{j}\left(\boldsymbol{\alpha}_{j}^{r}|\boldsymbol{\theta},\boldsymbol{y}_{j}\right).\label{eq:joint distribution of particles given parameters}
\end{equation}

Let $\boldsymbol{k}=\left(k_{1},...,k_{S}\right)$,
with each $k_{j}\in\left\{ 1,...,R\right\} $, $\boldsymbol{\alpha}_{1:S}^{\boldsymbol{k}}=\big(\boldsymbol{\alpha}_{1}^{k_{1}},...,\boldsymbol{\alpha}_{S}^{k_{S}}\big)$ be a vector of all selected individual random effects, and
$\boldsymbol{\alpha}_{1:S}^{\left(-\boldsymbol{k}\right)}=\big\{ \boldsymbol{\alpha}_{1}^{\left(-k_{1}\right)},...,\boldsymbol{\alpha}_{S}^{\left(-k_{S}\right)}\big\}$ is a collection of all particles excluding the selected individual random effects with $\boldsymbol{\alpha}_{j}^{\left(-k_{j}\right)}=\big(\boldsymbol{\alpha}_{j}^{1},...,\boldsymbol{\alpha}_{j}^{k_{j}-1}, \boldsymbol{\alpha}_{j}^{k_{j}+1},...,\boldsymbol{\alpha}_{j}^{R}\big)$. 

The augmented target density is defined as 
\begin{equation}
\widetilde{p}_{R}\left(\boldsymbol{\theta},\boldsymbol{\alpha}_{1:S}^{1:R},\boldsymbol{k}\right|\boldsymbol{y}):=\frac{p\left(\boldsymbol{\theta},\boldsymbol{\alpha}_{1:S}^{\boldsymbol{k}}\right|\boldsymbol{y})}{R^{S}}\frac{\psi\left(\boldsymbol{\alpha}_{1:S}^{1:R}\right|\boldsymbol{y},\boldsymbol{\theta})}{\prod_{j=1}^{S}m_{j}\big(\boldsymbol{\alpha}_{j}^{{k}_{j}}|\boldsymbol{\theta},\boldsymbol{y}_{j}\big)}.\label{eq:target density PMCMC}
\end{equation}
To understand the role of the index vector $\boldsymbol{k}=(k_{1},...,k_{S})$, we note that the density $p\left(\boldsymbol{\theta},\boldsymbol{\alpha}_{1:S}^{\boldsymbol{k}}\right|\boldsymbol{y})$ in \eqref{eq:target density PMCMC} indicates that of all the random effect replicates $\boldsymbol{\alpha}_{1:S}^{1:R}$ it is $\boldsymbol{\alpha}_{1:S}^{\boldsymbol{k}}=(\boldsymbol{\alpha}_{1}^{{k_1}},...,\boldsymbol{\alpha}_{S}^{{k_S}})$ that is generated from the posterior $p\left(\boldsymbol{\theta},\boldsymbol{\alpha}_{1:S}\right|\boldsymbol{y})$ (after the sampler has converged to the target distribution). 
\citet{Gunawan2017} show more formally that the marginal density of $\bs \theta$ and $\boldsymbol{\alpha}_{1:S}$
with respect to the joint density $\widetilde{p}_{R}\left(\boldsymbol{\theta},\boldsymbol{\alpha}_{1:S}^{1:R},\boldsymbol{k}\right|\boldsymbol{y})$
is $p\left(\boldsymbol{\theta},\boldsymbol{\alpha}_{1:S}\right|\boldsymbol{y})$
and give convergence results for the PMwG sampler.

\subsubsection*{The Conditional Monte Carlo (MC) Algorithm}

The conditional MC algorithm outlined in \algref{alg:conditional Monte-Carlo-Algorithm-1} is an important component of the PMwG sampler and updates $R-1$ particles simultaneously, while keeping the particle $\boldsymbol{\alpha}_{1:S}^{\boldsymbol{k}}$ from the posterior $p\left(\boldsymbol{\theta},\boldsymbol{\alpha}_{1:S}\right|\boldsymbol{y})$ fixed in $\widetilde{p}_{R}\left(\boldsymbol{\alpha}_{1:S}^{1:R}|\boldsymbol{\theta},\boldsymbol{y}\right)$. Hence, the density of all the particles that are generated by the MC algorithm conditional on $\left(\boldsymbol{\alpha}{}_{1:S}^{\boldsymbol{k}},\boldsymbol{\theta},\boldsymbol{y}\right)$ is, \[{\psi\left(\boldsymbol{\alpha}_{1:S}^{1:R}\right|\boldsymbol{y},\boldsymbol{\theta})}/{\prod_{j=1}^{S}m_{j}\big(\boldsymbol{\alpha}_{j}^{{k}_{j}}| \boldsymbol{\theta},\boldsymbol{y}_{j}\big)} \] which appears in the augmented target density in  \eqref{eq:target density PMCMC}. 

The following simple example illustrates the notation and the target distribution in \eqref{eq:target density PMCMC}. Suppose that there are $S=2$ individuals and $R=3$ particles for each individual. Let $\boldsymbol{\alpha}_{1}=\left(\boldsymbol{\alpha}_{1}^{1},\boldsymbol{\alpha}_{1}^{2},\boldsymbol{\alpha}_{1}^{3}\right)$ and $\boldsymbol{\alpha}_{2}=\left(\boldsymbol{\alpha}_{2}^{1},\boldsymbol{\alpha}_{2}^{2},\boldsymbol{\alpha}_{2}^{3}\right)$ be vectors of particles for subjects $1$ and $2$, respectively. If $k_{1}=2$ and $k_{2}=3$, then we define $\boldsymbol{\alpha}_{1:2}^{\boldsymbol{k}}=\left(\boldsymbol{\alpha}_{1}^{2},\boldsymbol{\alpha}_{2}^{3}\right)$ as the vector of selected individual random effects for subjects $1$ and $2$, $\boldsymbol{\alpha}_{1}^{\left(-k_{1}\right)}=\left(\boldsymbol{\alpha}_{1}^{1},\boldsymbol{\alpha}_{1}^{3}\right)$ is the collection of all particles excluding the selected random effects for subject $1$, and $\boldsymbol{\alpha}_{2}^{\left(-k_{2}\right)}=\left(\boldsymbol{\alpha}_{2}^{1},\boldsymbol{\alpha}_{2}^{2}\right)$ is the collection of all particles excluding the selected random effects for subject $2$. Then the particles $\boldsymbol{\alpha}_{1:2}^{\boldsymbol{k}}$ are from the posterior  $p\left(\boldsymbol{\theta},\boldsymbol{\alpha}_{1:2}\right|\boldsymbol{y})$ and the rest of the particles $\boldsymbol{\alpha}_{1:2}^{(-\boldsymbol{k})}$ from the proposal distributions $m_{j}\left(\boldsymbol{\alpha}_{j}|\boldsymbol{\theta},\boldsymbol{y}_{j}\right)$ for $j=1,2$. The index $\boldsymbol{k}$ indicates the $\boldsymbol{\alpha}$ generated from the posterior $p(\boldsymbol{\theta},\boldsymbol{\alpha}_{1:S}|\boldsymbol{y})$  and is an important element of the PMwG sampler described by Algorithm \ref{alg:PMMH+PG algorithm-LBA}.

%The conditional Monte Carlo algorithm generates all particles from the proposals except the particles $\boldsymbol{\alpha}_{1:2}^{\boldsymbol{k}}$ from the posterior.

\begin{algorithm}
\caption{Conditional MC Algorithm}\label{alg:conditional Monte-Carlo-Algorithm-1}

\begin{enumerate}
\item Fix $\boldsymbol{\alpha}{}_{1:S}^{1}=\boldsymbol{\alpha}{}_{1:S}^{\boldsymbol{k}}$.
\item For $j=1,..,S$

\begin{enumerate}
\item Sample $\boldsymbol{\alpha}{}_{j}^{r}$ from the proposal density $m_{j}\left(\boldsymbol{\alpha}_{j}|\boldsymbol{\theta},\boldsymbol{y}_{j}\right)$
for $r=2,...,R$.
\item Compute the importance weights $\widetilde{w}_{j}^{r}=\frac{p\left(\mathbf{y}_{j}|\boldsymbol{\alpha}{}_{j}^{r},\boldsymbol{\theta}\right)p\left(\boldsymbol{\alpha}{}_{j}^{r}|\boldsymbol{\theta}\right)}{m_{j}\left(\boldsymbol{\alpha}_{j}^{r}|\boldsymbol{\theta},\boldsymbol{y}_{j}\right)}$, and normalized weights $\widetilde{W}_{j}^{r}={\widetilde{w}_{j}^{r}}/{\sum_{k=1}^{R}\widetilde{w}_{j}^{k}}$,
for $r=1,...,R$.
\end{enumerate}
\end{enumerate}
Note that step 2 is easily parallelized  across the $r=1,...,R$ particles and across the $j=1,...,S$ subjects, or even both. There is no dependence between these parallel computations within each step.

\end{algorithm}

\subsubsection*{Particle Metropolis within Gibbs (PMwG) Sampling}
PMwG samples from the
augmented target density in \eqref{eq:target density PMCMC}, 
which means that when the PMwG sampler has converged it generates 
samples $(\boldsymbol{\theta},\boldsymbol{\alpha}_{1:S}^{\boldsymbol{k}})$ from 
$p\left(\boldsymbol{\theta},\boldsymbol{\alpha}_{1:S}\right|\boldsymbol{y})$.

\algref{alg:PMMH+PG algorithm-LBA} describes the PMwG sampling scheme for the hierarchical LBA model 
defined in Section \ref{sec:The-Linear-Ballistic accumulator}. The sampler starts at an initial set of
parameters $\boldsymbol{\theta}=(\bs\mu,\bs\Sigma)$
and random effects $\boldsymbol{\alpha}_{1:S}$. We now explain one iteration of the PMwG algorithm. Steps~(2a)--(2c) of the algorithm sample the group-level parameters of the LBA model using Gibbs steps conditional on the selected particles $\boldsymbol{\alpha}_{1:S}^{\boldsymbol{k}}$ from previous iteration. Step (3) is the conditional MC algorithm that generates $R-1$ new particles
while keeping the particles $\boldsymbol{\alpha}_{1:S}^{\boldsymbol{k}}$
fixed and setting the first set of particles $\boldsymbol{\alpha}_{1:S}^{1}=\boldsymbol{\alpha}_{1:S}^{\boldsymbol{k}}$. We now have a collection of particles $\boldsymbol{\alpha}^{1:R}_{1:S}=(\boldsymbol{\alpha}^{{1}}_{1},\boldsymbol{\alpha}^{2}_{1},...,\boldsymbol{\alpha}^{R}_{1},...,\boldsymbol{\alpha}^{1}_{S},\boldsymbol{\alpha}^{2}_{S},...,\boldsymbol{\alpha}^{R}_{S})$, where $\boldsymbol{\alpha}^{1}_{j}=\boldsymbol{\alpha}^{k_j}_{j}$ for $j=1,..,S$.
The conditional Monte Carlo gives the particles $\boldsymbol{\alpha}_{1:S}^{1:R}$
and the normalised weights $\widetilde{W}_{1:S}^{1:R}$. Step (4) samples
the new index vector $\boldsymbol{k}=\left(k_{1},...,k_{S}\right)$ with
probability given by \eqref{eq:probPMwG}, updates the selected particles
$\boldsymbol{\alpha}_{1:S}^{\boldsymbol{k}}=\left(\boldsymbol{\alpha}_{1}^{k_{1}},\boldsymbol{\alpha}_{2}^{k_{2}},...,\boldsymbol{\alpha}_{S}^{k_{S}}\right)$, and discards the rest of the particles $\boldsymbol{\alpha}_{1:S}^{\boldsymbol{(-k)}}$. 

%Step (3) is the conditional MC algorithm that
%generates the ${\boldsymbol\alpha}_{1:S}^{\left(-\boldsymbol{k}\right)}$,
%and Step 4 samples the index vector $\boldsymbol{k}=\left({k}_{1},...,{k}_{S}\right)$.
Note that Step 2 in \algref{alg:conditional Monte-Carlo-Algorithm-1}
can easily be parallelized for $r=1,...,R$ particles and for $j=1,...,S$ subjects, which is one of the main computational advantages of the PMwG approach. 

The PMwG sampler is applied in three stages to improve its effectiveness: burnin, adaptation, and sampling stages.
The burnin stage allows the Markov chain to move from its initial position, which is 
randomly drawn from the prior, to the typical set of the posterior, 
i.e. the region of greatest posterior concentration \citep{Betancourt2018}.
The adaptation stage draws samples from a reasonable approximation to the posterior
distribution and uses those samples to construct improved proposal distributions for the sampling stage.
These adapted proposal distributions are further adapted in the sampling stage 
and allow for very efficient sampling. \ref{tuningPMwG} discusses the practical implementation of the PMwG sampler.

\begin{algorithm}
\caption{PMwG Algorithm for the LBA Model}\label{alg:PMMH+PG algorithm-LBA}
\begin{enumerate}
\item Select initial values for $\bs \alpha_{1:S}$, $\boldsymbol{\theta}$ and set $\bs\alpha_j^{1}:=\bs\alpha_j^{k_j}$.
\item \begin{enumerate}
\item  Sample $\bs \mu|\boldsymbol{k},\bs{\alpha}_{1:S}^{\boldsymbol{k}},\boldsymbol{\theta}_{-\bs \mu},\boldsymbol{y}$
from $\textrm{N}\left(\ov{\bs \mu},\ov{\bs \Sigma}\right)$,
where $\ov{\bs \Sigma}=\left(S {\bs\Sigma}^{-1}+I\right)^{-1}$
and $\ov{\bs \mu}=\ov {\bs \Sigma}\left ( {\bs \Sigma}^{-1}\sum_{j=1}^{S}\boldsymbol{\alpha}_{j}^{k_j}\right)$.
\item Sample $\bs \Sigma|\boldsymbol{k},\boldsymbol{\alpha}_{1:S}^{\boldsymbol{k}},\boldsymbol{\theta}_{-\bs\Sigma},\boldsymbol{y}$
from $IW\left(k_\alpha, \bs B_\alpha\right)$, where $k_\alpha=\v+D_\alpha-1+S$
and $\bs B_\alpha=2\v\,\textrm{diag}\left(1/a_{1},...,1/a_{D}\right)+\sum_{j=1}^{S}
\big(\bs{\alpha}_{j}^{k_j}-\bs \mu
\big)\big(\bs{\alpha}_{j}^{k_j}-\bs \mu\big)^{\top}$.
\item Sample $a_{d}|\boldsymbol{k},\boldsymbol{\alpha}_{1:S}^{\boldsymbol{k}},\boldsymbol{\theta}_{-a_{d}},\boldsymbol{y}$
from $IG\left(\frac{\v+D_\alpha}{2},\v\left(\bs\Sigma^{-1}\right)_{dd}+\frac{1}{\A_{d}^{2}}\right)$
for $d=1,...,D_\alpha$.
\end{enumerate}
\item Sample $\boldsymbol{\alpha}_{1:S}^{\left(-\boldsymbol{k}\right)}\sim\widetilde{p}_{R}\left(\cdotp|\boldsymbol{k},
    \boldsymbol{\alpha}_{1:S}^{\boldsymbol{k}},\boldsymbol{\theta},\boldsymbol{y}\right)$
using \algref{alg:conditional Monte-Carlo-Algorithm-1}.

\item Sample the index vector $\boldsymbol{k}=\left( k_{1},..., k_{S}\right)$
with probability given by

\begin{equation}
\widetilde{p}_{R}\left(k_{1}=l_{1},...,k_{S}=l_{S}|\boldsymbol{\theta},\boldsymbol{\alpha}_{1:S}^{1:R},\boldsymbol{y}\right)=\prod_{j=1}^{S}\widetilde{W}_{j}^{l_{j}}.\label{eq:probPMwG}
\end{equation}

\item Repeat steps 2 to 4 for the required number of iterations.

\end{enumerate}
\end{algorithm}

\subsection{\Dt_SMC{} for a random effects model (DT-SMC)
\label{subsec:Annealing-Importance-Sampling}}

The \dt_SMC{} method we use is an alternative approach to all MCMC methods for obtaining samples from the posterior density; it
builds on the SMC algorithm
of \citet{DelMoral:2006} and \citet{Duan2015} by propagating a particle cloud $\big(\boldsymbol{\theta}_{1:M}^{\left(p\right)},\boldsymbol{\alpha}_{1:M}^{\left(p\right)},W_{1:M}^{\left(p\right)}\big)$
through a sequence of tempered target densities $\xi_{a_{p}}\left(\boldsymbol{\theta},\boldsymbol{\alpha}|\boldsymbol{y}\right)$,
for $p=0,...,P$, to the posterior density $p\left(\boldsymbol{\theta},\boldsymbol{\alpha}|\boldsymbol{y}\right)$. 
The sequence of tempered densities is defined as
\begin{equation}\label{eq: annealed densities}
\xi_{a_{p}}\left(\boldsymbol{\theta},\boldsymbol{\alpha}|\boldsymbol{y}\right) :=\eta_{a_{p}}\left(\boldsymbol{\theta},\boldsymbol{\alpha}|\boldsymbol{y}\right)/Z_{a_{p}},\;\;\;\;\textrm{ with }\;\;\;\;Z_{a_{p}} =\int\eta_{a_{p}}\left(\boldsymbol{\theta},\boldsymbol{\alpha}|\boldsymbol{y}\right)d\boldsymbol{\theta}d\boldsymbol{\alpha}; 
\end{equation}
with $ 0=a_{0}<a_{1}<...<a_{P}=1$ and 
\[
\eta_{a_{p}}\left(\boldsymbol{\theta},
\boldsymbol{\alpha}|\boldsymbol{y}\right):=\left(p_{0}\left(\boldsymbol{\theta},\boldsymbol{\alpha}\right)\right)^{1-a_{p}}\left(p\left(\boldsymbol{y}|
\boldsymbol{\theta},\boldsymbol{\alpha}\right)p\left(\boldsymbol{\alpha}|\boldsymbol{\theta}\right)p\left(\boldsymbol{\theta}\right)\right)^{a_{p}}.\]
The \dt_SMC algorithm produces the $M$ triples  $\{\big(\boldsymbol{\theta}_{m}^{\left(P\right)},\boldsymbol{\alpha}_{m}^{\left(P\right)},W_{m}^{\left(P\right)}\big), \\ m=1,\dots,M\},$ which approximate the posterior density $p\left(\boldsymbol{\theta},\boldsymbol{\alpha}|\boldsymbol{y}\right)$.

We take $p_{0}\left(\boldsymbol{\theta},\boldsymbol{\alpha}\right):=p\left(\boldsymbol{\theta}\right)
p\left(\boldsymbol{\alpha}|\boldsymbol{\theta}\right)$, because in the 
current LBA model, it is both easy to generate from and evaluate the prior densities $p\left(\boldsymbol{\theta}\right)$ and $p\left(\boldsymbol{\alpha}|\boldsymbol{\theta}\right)$. With this choice, 
\[
\eta_{a_{p}}\left(\boldsymbol{\theta},\boldsymbol{\alpha}|\boldsymbol{y}\right)=p\left(\boldsymbol{y}|\boldsymbol{\theta},\boldsymbol{\alpha}\right)^{a_{p}}p\left(\boldsymbol{\alpha}|\boldsymbol{\theta}\right)p\left(\boldsymbol{\theta}\right).
\]

At the initial temperature, the particle
cloud $\big\{ \boldsymbol{\theta}_{1:M}^{\left(0\right)},\boldsymbol{\alpha}_{1:M}^{\left(0\right)},W_{1:M}^{\left(0\right)}\big\} $
is obtained by sampling $\big\{ \boldsymbol{\theta}_{1:M}^{\left(0\right)},\boldsymbol{\alpha}_{1:M}^{\left(0\right)}\big\} $
from  $p_{0}\left(\boldsymbol{\alpha},\boldsymbol{\theta}\right)$,
and giving all particles equal weight, $W_{1:M}^{\left(0\right)}=1/M$.
The particle cloud $\big\{ \boldsymbol{\theta}_{1:M}^{\left(p-1\right)},\boldsymbol{\alpha}_{1:M}^{\left(p-1\right)},W_{1:M}^{\left(p-1\right)}\big\} $
at iteration $p-1$ is an estimate of $\xi_{a_{p-1}}\left(\boldsymbol{\theta},\boldsymbol{\alpha}|\boldsymbol{y}\right)$.
The transition from the particle cloud estimate of $\xi_{a_{p-1}}\left(\boldsymbol{\theta},\boldsymbol{\alpha}|\boldsymbol{y}\right)$
to the particle cloud estimate of  $\xi_{a_{p}}\left(\boldsymbol{\theta},\boldsymbol{\alpha}|\boldsymbol{y}\right)$
is implemented by first reweighting to obtain
the updated weights $W_{1:M}^{\left(p\right)}= {w_{1:M}^{(p)}}/{\sum_{j=1}^{M}w_{j}^{(p)}}$, where
\begin{equation}\label{eq:updated weights}
w_{m}^{\left(p\right)}=W_{m}^{\left(p-1\right)}\frac{\eta_{a_{p}}
\big(\boldsymbol{\theta}_{m}^{(p-1)},\boldsymbol{\alpha}_{m}^{(p-1)}|\boldsymbol{y}\big)}{\eta_{a_{p-1}}\big(\boldsymbol{\theta}_{m}^{(p-1)},\boldsymbol{\alpha}_{m}^{(p-1)}|\boldsymbol{y}\big)}
=W_{m}^{\left(p-1\right)}p\left(\boldsymbol{y}|\boldsymbol{\theta}_{m}^{(p-1)},\boldsymbol{\alpha}_{m}^{(p-1)}\right)^{a_{p}-a_{p-1}}.
\end{equation}
We now follow \citet{DelMoral2012} and select the next value of $a_{p}$
to target a pre-defined effective sample size, $\textrm{ESS}_{T}$\footnote{ESS measures variability in the weights, and is defined as  $\textrm{ESS}^{-1}=\sum_{i=1}^{M}\left(W_{i}^{\left(p-1\right)}\right)^{2}$, and varies between 1 and $M$. A low value of ESS indicates that
the weights are concentrated on only a few particles.}. We do so by evaluating the ESS over a grid of points $a_{1:G,p}$ of
potential $a_{p}$ values and select as $a_{p}$ the value of $a_{j,p}$
whose ESS is the closest to $\textrm{ESS}_{T}$. After reweighting, the effective sample size (ESS) is close to $\textrm{ESS}_{T}$. To eliminate particles with low weight and replicate particles with larger weights,
$\big\{ \boldsymbol{\theta}_{1:M}^{\left(p-1\right)},\boldsymbol{\alpha}_{1:M}^{\left(p-1\right)}\big\} $
are resampled with probabilities given by their normalised
weights $W_{1:M}^{(p)}$.

To improve the approximation of the particle cloud to $\xi_{a_p}$, we carry out $L$ Markov move steps
for each particle, using a Markov kernel $K_{\xi_{a_{p}}}$ that has
$\xi_{a_{p}}$ as its invariant density. This Markov move step increases particle diversity, and in particular 
makes identical particles (which are produced during re-sampling) different from each other. 

%Repeated application of this procedure can reduce the particle diversity so that the
%particle cloud at the $p^{th}$ iteration may not be a good approximation
% to the $p$th tempered density $\xi_{a_p}$. 
%To improve the approximation, we carry out $L$ Markov move steps
%for each particle, using a Markov kernel $K_{\xi_{a_{p}}}$ that has
%$\xi_{a_{p}}$ as its invariant density. These moves serve to make identical particles (which are produced during re-sampling) %different from each other. 

The Markov kernel $K_{\xi_{a_{p}}}$ is constructed based on the PMwG algorithm.
The augmented tempered target density at SMC step $p$ is defined as
\begin{equation}
\widetilde{\xi}_{a_{p}}\left(\boldsymbol{\theta},\boldsymbol{\alpha}_{1:S}^{1:R},\boldsymbol{k}|\boldsymbol{y}\right)
:=\frac{\xi_{a_{p}}\left(\boldsymbol{\theta},\boldsymbol{\alpha}_{1:S}^{\boldsymbol{k}}|\boldsymbol{y}\right)}{R^{S}}
\frac{\psi\left(\boldsymbol{\alpha}_{1:S}^{1:R}|\boldsymbol{\theta},\boldsymbol{y}\right)}{\prod_{j=1}^{S}m_{j}\big(\boldsymbol{\alpha}_{j}^{{k}_{j}}|
\boldsymbol{\theta},\boldsymbol{y}_{j}\big)},\label{eq:AISrandeffect-1}
\end{equation}
where \eqref{eq:joint distribution of particles given parameters} gives $\psi\left(\boldsymbol{\alpha}_{1:S}^{1:R}|{\boldsymbol{\theta}},\boldsymbol{y}\right)$; 
we note that it is possible to use different proposal densities $m_{j}^{(p)}(\cdot)$ for different SMC steps $p$.
Using the same derivation as in \citet{Gunawan2017}, we can show that
the marginal density of $\boldsymbol{\theta}$ and $\boldsymbol{\alpha}_{1:S}$ w.r.t. $\widetilde{\xi}_{a_{p}}\left(\boldsymbol{\theta},\boldsymbol{\alpha}_{1:S}^{1:R},\boldsymbol{k}|\boldsymbol{y}\right)$ is $\xi_{a_{p}}\left(\boldsymbol{\theta},\boldsymbol{\alpha}_{1:S}|\boldsymbol{y}\right)$.
The augmented tempered density involves the term
$\psi\left(\boldsymbol{\alpha}_{1:S}^{1:R}|{\boldsymbol{\theta}},\boldsymbol{y}\right)/{\prod_{j=1}^{S}m_{j}\big(\boldsymbol{\alpha}_{j}^{\boldsymbol{k}_{j}}|\boldsymbol{\theta},\boldsymbol{y}_{j}\big)}$,
which is the density under $\widetilde{\xi}_{a_{p}}$ of all particles
that are generated by the MC algorithm conditional on $\left(\boldsymbol{\alpha}_{1:S}^{\bs k},\boldsymbol{\theta},\boldsymbol{y}\right)$.
The conditional MC algorithm is
similar to the one given in Algorithm \ref{alg:conditional Monte-Carlo-Algorithm-1},
except that the \dt_SMC{}  version adopts the tempered conditional density $p\left(\boldsymbol{y}|\boldsymbol{\theta},\boldsymbol{\alpha}\right)^{a_{p}}$
instead of $p\left(\boldsymbol{y}|\boldsymbol{\theta},\boldsymbol{\alpha}\right)$.
Therefore, the Markov move step is based on the PMwG sampling scheme
in \algref{alg:PMMH+PG algorithm-LBA}, except that instead of $\widetilde{p}_{R}$,
we have augmented tempered target densities $\widetilde{\xi}_{a_{p}}$.

\algref{alg:Generic-AISIL-Algorithm} describes the \dt_SMC{} algorithm. 
Steps~(1), (2a)-(2d) are standard and apply to any model with slight modification.
Step~(2e) performs $M$ parallel PMwG algorithm (\algref{alg:PMMH+PG algorithm-LBA}) $L$ times for each temperature, except that in Step 2 of that algorithm we sample $\boldsymbol{\alpha}_{1:S}^{\left(-\boldsymbol{k}\right)}\sim\widetilde{\xi}_{a_{p}}\left(\cdotp|\boldsymbol{k}, \boldsymbol{\alpha}_{1:S}^{\boldsymbol{k}},\boldsymbol{\theta},\boldsymbol{y}\right)$ using the conditional MC procedure in \algref{alg:conditional Monte-Carlo-Algorithm-1} with the likelihood $p\left(\boldsymbol{y}|\boldsymbol{\theta},\boldsymbol{\alpha}\right)$ replaced by $p\left(\boldsymbol{y}|\boldsymbol{\theta},\boldsymbol{\alpha}\right)^{a_{p}}$. This is an attractive and important feature as it allows the algorithm to fully use the computational power of modern graphical processing units (GPUs) with thousands of parallel cores.
\ref{tuningAISIL} discusses the tuning parameters and the proposal densities in the \dt_SMC{} algorithm.

%Step~(2e) performs $L$ Markov moves based on the PMwG algorithm (\algref{alg:PMMH+PG algorithm-LBA}),
%except that in Step 2 of that algorithm we sample
%$\boldsymbol{\alpha}_{1:S}^{\left(-\boldsymbol{k}\right)}\sim\widetilde{\xi}_{a_{p}}\left(\cdotp|\boldsymbol{k},
%\boldsymbol{\alpha}_{1:S}^{\boldsymbol{k}},\boldsymbol{\theta},\boldsymbol{y}\right)$ using the conditional MC procedure in \algref{alg:conditional Monte-Carlo-Algorithm-1}
%with the likelihood $p\left(\boldsymbol{y}|\boldsymbol{\theta},\boldsymbol{\alpha}\right)$ replaced by $p\left(\boldsymbol{y}|\boldsymbol{\theta},\boldsymbol{\alpha}\right)^{a_{p}}$. Thus, the DT-SMC algorithm entails running $M$ parallel PMwG algorithm $L$ times for each temperature. \ref{tuningAISIL} discusses the tuning parameters and the proposal densities in the \dt_SMC{} algorithm.

\begin{algorithm}
\caption{The \dt_SMC{} algorithm} \label{alg:Generic-AISIL-Algorithm}
\begin{enumerate}
\item Set $p=0$ and generate $\big\{ \boldsymbol{\theta}_{1:M}^{\left(0\right)},\boldsymbol{\alpha}_{1:M}^{\left(0\right)}\big\} $
from $p_{0}\left(\boldsymbol{\alpha},\boldsymbol{\theta}|\boldsymbol{y}\right)$,
and give them equal weight $W_{m}^{\left(0\right)}=1/M$,
for $m=1,...,M$.
\item While the tempering sequence $a_{p}<1$ do

\begin{enumerate}
\item Set $p\leftarrow p+1$.
\item Find $a_{p}$ adaptively by searching across a grid of $a_{p}$ to
maintain effective sample size near some constant $\textrm{ESS}_{T}$.
\item Compute new normalised weights $W_{1:M}^{\left(p\right)}= {w_{1:M}^{(p)}}/{\sum_{j=1}^{M}w_{j}^{(p)}}$ with unnormalized weights in \eqref{eq:updated weights}.
\item Resample $\big(\boldsymbol{\theta}_{m}^{\left(p-1\right)},\boldsymbol{\alpha}_{m}^{\left(p-1\right)}\big)$
using the weights $W_{1:M}^{\left(p\right)}$ to obtain\\ $\big(\boldsymbol{\theta}_{1:M}^{\left(p\right)},\boldsymbol{\alpha}_{1:M}^{\left(p\right)}\big)$.
\item Make $L$ Markov moves

\begin{enumerate}
\item Let $K_{a_{p}}\left(\left(\boldsymbol{\theta},\boldsymbol{\alpha}\right),\cdot\right)$
be a Markov kernel having invariant density $\xi_{a_{p}}\left(\boldsymbol{\theta},\boldsymbol{\alpha}|\boldsymbol{y}\right)$.
For $m=1,...,M$, move each $\big(\boldsymbol{\theta}_{m}^{\left(p\right)},\boldsymbol{\alpha}_{m}^{\left(p\right)}\big)$
$L$ times using the Markov kernel $K_{a_{p}}$ to obtain $\big(\widetilde{\boldsymbol{\theta}}_{m},\widetilde{\boldsymbol{\alpha}}_{m}\big)$.  The Markov move step is based on the PMwG
in Algorithm \ref{alg:PMMH+PG algorithm-LBA}, except that instead of $\widetilde{p}_{R}$,
we have augmented tempered target densities $\widetilde{\xi}_{a_{p}}$.
\item Set $\big(\boldsymbol{\theta}_{1:M}^{\left(p\right)},\boldsymbol{\alpha}_{1:M}^{\left(p\right)}\big)\leftarrow\big(\widetilde{\boldsymbol{\theta}}_{1:M},\widetilde{\boldsymbol{\alpha}}_{1:M}\big)$
and set $W_{1:M}^{\left(p\right)}=1/M$.
\end{enumerate}
\end{enumerate}
\end{enumerate}
\end{algorithm}

\subsection{Estimating the Marginal Likelihood\label{MarginalLikelihoodEstimation}}
This section shows show how to estimate the marginal likelihood  with negligible post-processing cost from the intermediate outputs of \dt_SMC{} using both the standard method \citep{DelMoral:2006} as well as thermodynamic integration (TI). 
%\ref{MarginalLikelihoodPMwG} discusses the estimation of marginal likelihood from the PMwG output. 

\subsection*{Standard \dt_SMC estimation of the marginal likelihood} 

The marginal likelihood 
\[
p\left(\boldsymbol{y}\right)=\prod_{p=1}^{P}\frac{Z_{a_{p}}}{Z_{a_{p-1}}}\;\;\textrm{with}\;\;\frac{Z_{a_{p}}}{Z_{a_{p-1}}}=\int\left(\frac{\eta_{a_{p}}\left(\boldsymbol{\theta},\boldsymbol{\alpha}|\boldsymbol{y}\right)}{\eta_{a_{p-1}}\left(\boldsymbol{\theta},\boldsymbol{\alpha}|\boldsymbol{y}\right)}\right)\widetilde{\xi}_{a_{p-1}}\left(\boldsymbol{\theta},\boldsymbol{\alpha}|\boldsymbol{y}\right)d\boldsymbol{\theta}d\boldsymbol{\alpha}.
\]
since   $Z_{a_{0}}=1$ and $p\left(\boldsymbol{y}\right)=Z_{a_{P}}$. 
The particle cloud $\left(\boldsymbol{\theta}_{1:M}^{\left(p-1\right)},
\boldsymbol{\alpha}_{1:M}^{\left(p-1\right)},W_{1:M}^{\left(p-1\right)}\right)$
approximates $\widetilde{\xi}_{a_{p-1}}\left(\boldsymbol{\theta},\boldsymbol{\alpha}|\boldsymbol{y}\right)$,
so that the ratio $Z_{a_p}/Z_{a_{p-1}} $ is estimated by
$\widehat{Z_{a_{p}}/Z_{a_{p-1}}}=\sum_{m=1}^{M}w_{m}^{\left(p\right)}$,
giving the marginal likelihood estimate
\[
\widehat{p}\left(\boldsymbol{y}\right)=\prod_{p=1}^{P}\widehat{\frac{Z_{a_{p}}}{Z_{a_{p-1}}}}.
\]

\subsection*{Thermodynamic Integration Estimators from \dt_SMC Output}
\citet{gelman1998simulating} and \citet{friel2008marginal} show how to compute the marginal likelihood
of the data given the model using ideas from thermodynamic integration
or path sampling, which relies on sampling from the posterior at different
temperatures (called ``power posteriors'' or ``tempered posteriors''). 

The thermodynamic identity \citep{friel2008marginal,Friel:2014} is
\begin{eqnarray}
\log p\left(y\right) & = & \int_{0}^{1}\E_{\xi_{a_{p}}}\left(\log\left\{ p\left(\boldsymbol{y}|\boldsymbol{\theta},\boldsymbol{\alpha}\right)\right\} \right)da_{p}.\label{eq:thermodynamicIdentity}
\end{eqnarray}
The log of the marginal likelihood is the integral of $E_{\xi_{a_{p}}}\left(\log\left\{ p\left(\boldsymbol{y}|\boldsymbol{\theta},\boldsymbol{\alpha}\right)\right\} \right)$
over the tempering sequence $a_{p}$, where $a_{p}$ moves from $0$
to $1$. The derivation of the thermodynamic identity is 
in \citet{friel2008marginal} and \citet{Friel:2014}. For each value of the tempering
sequence $a_{p}$, a sample from $\xi_{a_{p}}\left(\boldsymbol{\theta},\boldsymbol{\alpha}|\boldsymbol{y}\right)$
can be used to estimate $E_{\xi_{a_{p}}}\left(\log\left\{ p\left(\boldsymbol{y}|\boldsymbol{\theta},\boldsymbol{\alpha}\right)\right\} \right)$. \ref{DetailThermodynamicIntegration} describes first and second order quadrature approximations to the integral in \eqref{eq:thermodynamicIdentity}, labelled in the results below as $TI_{1}$ and $TI_{2}$, respectively. We present two approximations as $TI_{2}$ corrects for bias that can be introduced in the method of estimating $TI_{1}$. 

There are some advantages in estimating the marginal likelihood by thermodynamic integration 
using \dt_SMC{} output: (i)~\dt_SMC produces the $M$ triples
$\left(\boldsymbol{\theta}_{1:M}^{\left(p\right)},\boldsymbol{\alpha}_{1:M}^{\left(p\right)},W_{1:M}^{\left(p\right)}\right)$
for each value of the tempering sequence $a_{p}$, $p=0,...,P$;  this means that estimating the marginal likelihood via
thermodynamic integration does not incur any extra computational cost -- marginal likelihood can be calculated by TI directly from the \dt_SMC{} outputs; 
(ii)~assessing the convergence of \dt_SMC{} is much less of an issue than that of MCMC methods; 
and \dt_SMC{} is easily parallelizable for each $a_p$ in the tempering sequence; (iii)~the
number of tempering steps and the tempering sequence are chosen adaptively
to target a pre-defined effective sample size of the SMC samples.

\subsection{Estimating Marginal Likelihood using the PMwG output}
\ref{MarginalLikelihoodPMwG} discusses the estimation of marginal likelihood using the PMwG output. 

\newcommand{\muab}[3]{{\bs {#1}}_{{{#2}^{#3}}}}
\newcommand{\sigmab}[1]{{\bs \Sigma}_{{#1}}}

\section{Illustrative Applications}\label{sec:Simulation-Study-and real applications}

\subsection{Application to Simulated Data\label{subsec:Simulation-Study}}
This section applies the PMwG and \dt_SMC{} methods to fit the hierarchical LBA model specified in 
Section~\ref{sec:The-Linear-Ballistic accumulator}; the data simulated from an LBA model 
mimicking the conditions inspired by the experiment of \citet{Forstmann2008}. The  three generated datasets have 
$S=19$, $50$, and $100$ subjects, and $N = 1000$ trials ($N_{z}\approx333$ trials in each condition) for each subject for each value of $S$; 
the generated data have nonzero correlations between the random effects which reflects both the improved parametric specification of the LBA model as well as  plausible psychological assumptions about individual differences. To generate the simulated data, we used group-level parameters which matched those estimated for the real data reported by \citet{Forstmann2008}. This defined a multivariate normal distribution for the log-transformed random effects. The random effects for the simulated data sets were sampled randomly from this multivariate normal.

%We used 500 iterations for burnin and another 500 for adaptation. 
A total of $10,000$ draws were 
obtained using the PMwG sampler for the subsequent analysis of the posterior distribution. 
See \ref{tuningPMwG} for a discussion on the tuning parameters, the proposal 
densities, and the implementation of the PMwG sampler. 
We assess the performance of the sampler using the ``integrated autocorrelated time'' $\left(\textrm{IACT}\right)$, which
measures the inefficiency of the sampling scheme 
in terms of the multiple of its iterates that are required to obtain the same variance as an independent sampling scheme, e.g. if $\textrm{IACT}=10$, then we need 10 times as many iterates as an independent scheme, so a larger value indicates poorer performance. 
The IACT of a scalar parameter $\omega$ is defined as \citep{Chib1996}
\[
\textrm{IACT}_{\omega}:=1+2\sum_{t=1}^{\infty}\rho_{\omega}\left(t\right),
\]
where $\rho_{\omega}\left(t\right)$ is the lag-$t$ autocorrelation of the iterates
of $\omega$ for the underlying stationary Markov chain. 
Table~\ref{tab:Inefficiency-Factors-of LBA parameters simulation} reports the estimated IACT  
values for the group level parameters from the PMwG sampler, with the estimates obtained using the CODA R package of \citet{Plummer2006}\footnote{It does not make sense to calculate IACT for the \dt_SMC{} sampler, because it is not based on MCMC.}.  The results suggest that the PMwG sampler is efficient because all its IACT values are small. Similar conclusions can be drawn from the IACT values for the random effects, as shown in the online supplement at \url{osf.io/5b4w3}.

\begin{table}[H]
\caption{Estimated Inefficiency Factors (IACT) of the LBA parameters using the PMwG
method with $S=19,50,100$ subjects and $N=1,000$ in the simulation study.
The order of the random effect parameters in the covariance matrix $\Sigma$
is ${b^{\left(1\right)}}$, ${b^{\left(2\right)}}$,
${b^{\left(3\right)}}$, ${A}$, ${v^{\left(1\right)}}$,
${v^{\left(2\right)}}$, ${\tau}$.\label{tab:Inefficiency-Factors-of LBA parameters simulation}}
\centering{}%
\begin{tabular}{cccccccc}
\hline 
Param & $S=19$ & $S=50$ & $S=100$ & Param & $S=19$ & $S=50$ & $S=100$\tabularnewline
\hline 
$\boldsymbol{\mu}_1$ & $1.29$ & $1.15$ & $1.16$ & $\boldsymbol{\Sigma}_{11}$ & $1.89$ & $1.32$ & $1.29$\tabularnewline
$\boldsymbol{\mu}_2$ & $1.23$ & $1.15$ & $1.14$ & $\boldsymbol{\Sigma}_{22}$ & $1.79$ & $1.35$ & $1.21$\tabularnewline
$\boldsymbol{\mu}_3$ & $1.12$ & $1.11$ & $1.08$ & $\boldsymbol{\Sigma}_{33}$ & $1.47$ & $1.25$ & $1.16$\tabularnewline
$\boldsymbol{\mu}_4$ & $1.37$ & $1.88$ & $1.60$ & $\boldsymbol{\Sigma}_{44}$ & $1.99$ & $1.80$ & $1.63$\tabularnewline
$\boldsymbol{\mu}_5$ & $1.14$ & $1.10$ & $1.18$ & $\boldsymbol{\Sigma}_{55}$ & $1.72$ & $1.36$ & $1.49$\tabularnewline
$\boldsymbol{\mu}_6$ & $1.21$ & $1.16$ & $1.19$ & $\boldsymbol{\Sigma}_{66}$ & $2.22$ & $1.55$ & $1.41$\tabularnewline
$\boldsymbol{\mu}_7$ & $1.35$ & $2.05$ & $1.65$ & $\boldsymbol{\Sigma}_{77}$ & $2.46$ & $3.60$ & $2.31$\tabularnewline
\hline 
$\boldsymbol{\Sigma}_{12}$ & $1.88$ & $1.34$ & $1.26$ & $\boldsymbol{\Sigma}_{34}$ & $1.82$ & $1.45$ & $1.44$\tabularnewline
$\boldsymbol{\Sigma}_{13}$ & $1.81$ & $1.29$ & $1.25$ & $\boldsymbol{\Sigma}_{35}$ & $1.33$ & $1.25$ & $1.34$\tabularnewline
$\boldsymbol{\Sigma}_{14}$ & $1.85$ & $1.53$ & $1.38$ & $\boldsymbol{\Sigma}_{36}$ & $1.53$ & $1.48$ & $1.28$\tabularnewline
$\boldsymbol{\Sigma}_{15}$ & $1.59$ & $1.30$ & $1.39$ & $\boldsymbol{\Sigma}_{37}$ & $1.78$ & $2.12$ & $1.45$\tabularnewline
$\boldsymbol{\Sigma}_{16}$ & $1.61$ & $1.47$ & $1.29$ & $\boldsymbol{\Sigma}_{45}$ & $1.72$ & $1.47$ & $1.78$\tabularnewline
$\boldsymbol{\Sigma}_{17}$ & $2.08$ & $2.25$ & $1.50$ & $\boldsymbol{\Sigma}_{46}$ & $1.57$ & $1.54$ & $1.40$\tabularnewline
$\boldsymbol{\Sigma}_{23}$ & $1.72$ & $1.30$ & $1.22$ & $\boldsymbol{\Sigma}_{47}$ & $1.99$ & $1.37$ & $1.43$\tabularnewline
$\boldsymbol{\Sigma}_{24}$ & $1.83$ & $1.50$ & $1.37$ & $\boldsymbol{\Sigma}_{56}$ & $1.52$ & $1.51$ & $1.35$\tabularnewline
$\boldsymbol{\Sigma}_{25}$ & $1.75$ & $1.30$ & $1.36$ & $\boldsymbol{\Sigma}_{57}$ & $1.61$ & $1.61$ & $1.58$\tabularnewline
$\boldsymbol{\Sigma}_{26}$ & $1.56$ & $1.51$ & $1.28$ & $\boldsymbol{\Sigma}_{67}$ & $1.75$ & $1.44$ & $1.33$\tabularnewline
$\boldsymbol{\Sigma}_{27}$ & $2.02$ & $2.26$ & $1.47$ &  &  &  & \tabularnewline
\hline 
\end{tabular}
\end{table}

We used 10 independent runs with $M=250$ samples each for the \dt_SMC{} method to generate $2,500$ samples of the LBA
individual random effects and parameters. The independent samplers mean that $10$ independent estimates of the marginal likelihood are also obtained, which allows a rough estimate of the sampling variability in the marginal likelihood. Sampling error in the marginal likelihood is important in inference, but often overlooked. 
See \ref{tuningAISIL} for a discussion of the tuning parameters, the proposal 
densities, and the implementation of the \dt_SMC{} sampler.

Table \ref{tab:computation times PMwG and AISIL-RE simulations} shows the wall-clock computation time to run both the PMwG 
and the \dt_SMC{} methods which are based on a Matlab implementation running on 28 CPU-cores. The running time for PMwG includes the time taken for all three stages. 
The table shows that PMwG is much faster than \dt_SMC{}. In general, the PMwG method can be used with minimal computational
resources, e.g. a personal computer;
however, \dt_SMC{} is easier to parallelize than PMwG, so it is likely to be faster than PMwG if 
there is access to many more CPU cores and the model or data are large. 
The optimal number of CPU cores required for \dt_SMC{} is equal to the number of SMC samples $M$, which means the properties of the sampler can be easily tuned to provide maximum parallel efficiency on a large range of hardware.

\begin{table}[H]
\caption{Computation time (in minutes) for PMwG and \dt_SMC{} for $N$ trials and $S$ subjects running on 28 CPU-cores. \label{tab:computation times PMwG and AISIL-RE simulations}}

\centering{}%
\begin{tabular}{cccc}
\hline 
S & N & PMwG & \dt_SMC{} \tabularnewline
\hline 
19 & 1,000 & 36 & 180\tabularnewline
50 & 1,000 & 66 & 688\tabularnewline
100 & 1,000 & 193 & 1,810\tabularnewline
\hline 
\end{tabular}
\end{table}

\begin{figure}[h]
\centering{}\includegraphics[width=13cm,height=10cm]{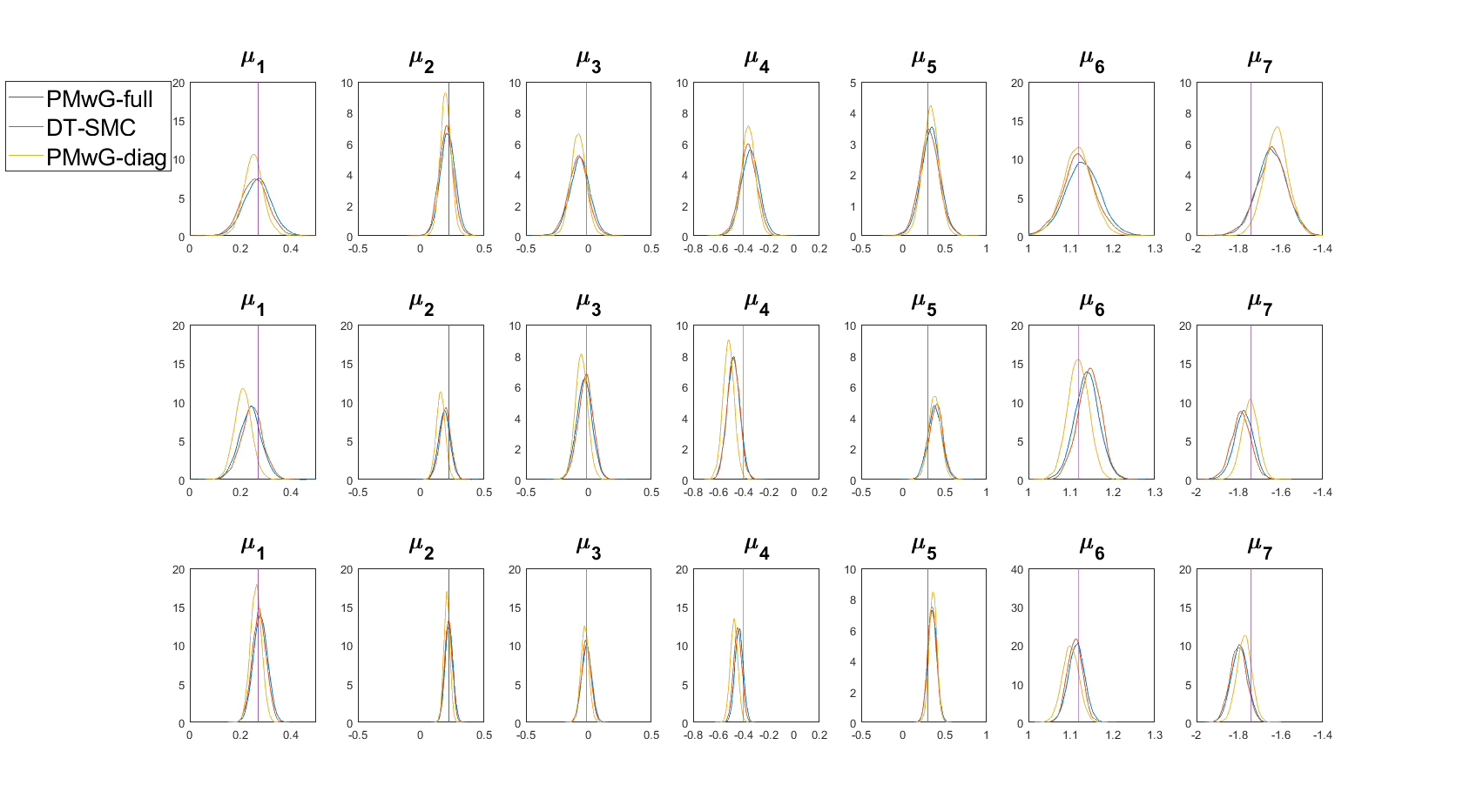}
\caption{Kernel density estimates of the LBA group level mean parameters with $S=19$ subjects 
(top), $S=50$ (middle), and $S=100$ (bottom) and $N=1,000$ in the simulation study. Colours represent the three sampling algorithms: PMwG (blue); \dt_SMC{} (red); and PMwG with a diagonal covariance matrix (no between-parameter correlations in the prior, yellow). The vertical lines show the true (data generating) values. \label{fig:kerneldensitysimthetaS19T1000}}
\end{figure}

\begin{figure}[h]
\centering{}\includegraphics[width=13cm,height=10cm]{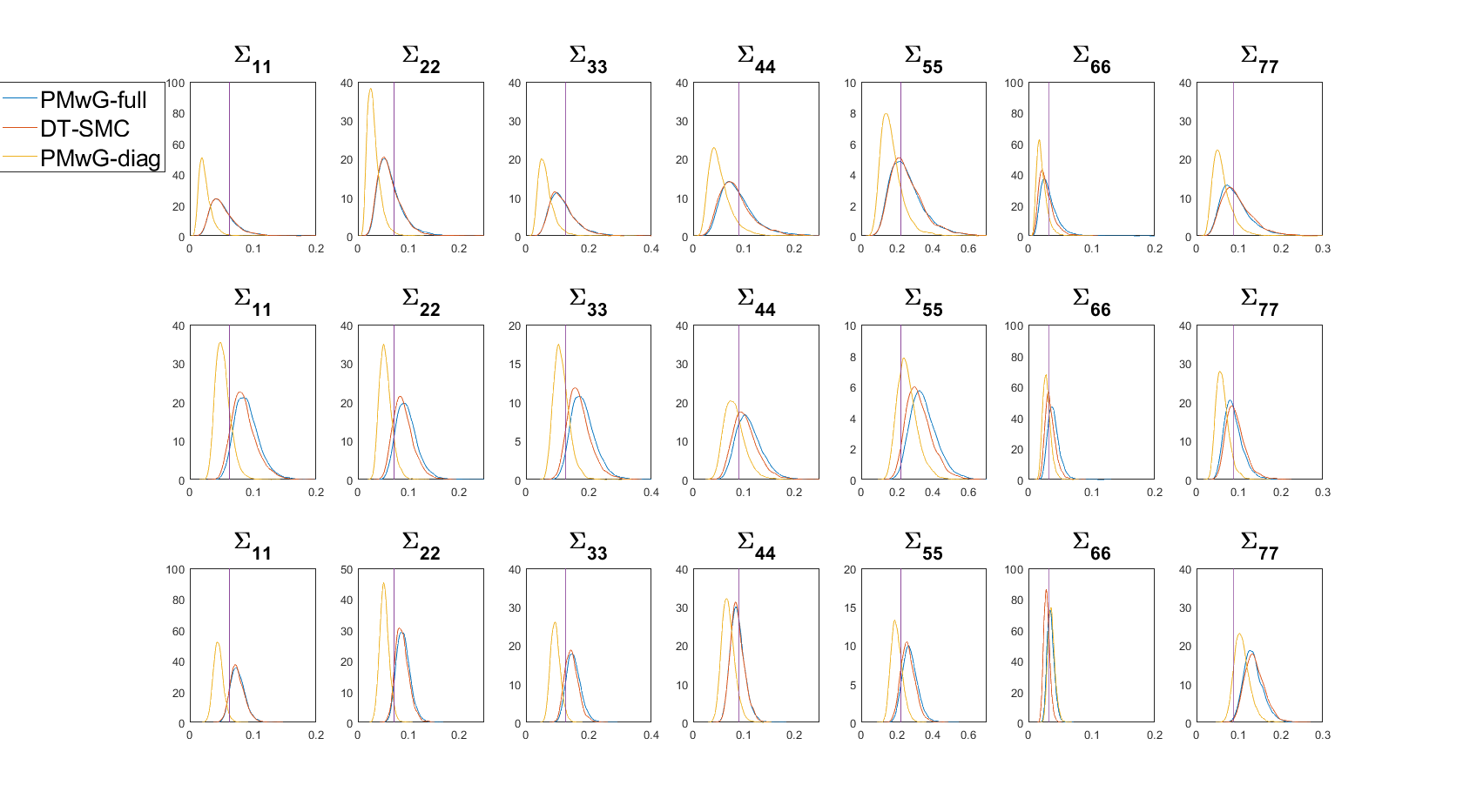}
\caption{Kernel density estimates of the LBA group level variance parameters with $S=19$ subjects (top), $S=50$ (middle), and $S=100$ (bottom) and $N=1,000$ in the simulation study. Colours represent the three sampling algorithms: PMwG (blue);
\dt_SMC{} (red); and PMwG with a diagonal covariance matrix (no between-parameter correlations, yellow). The vertical lines show the true (data generating) values. \label{fig:kerneldensitysimsigmaS19T1000}}
\end{figure}

Figures \ref{fig:kerneldensitysimthetaS19T1000} and \ref{fig:kerneldensitysimsigmaS19T1000} 
summarize the results for the simulated data. Figure~\ref{fig:kerneldensitysimthetaS19T1000}
plots the posterior distributions estimated by PMwG (in blue)  
and \dt_SMC{} (in red)
for the population mean parameters ($\mu$), with the vertical lines showing
the true parameter values.  Figure~\ref{fig:kerneldensitysimsigmaS19T1000} shows similar plots for 
the variances (the diagonal elements of $\Sigma$) of the random effects. 
In both figures, the top, middle and bottom panels correspond to $S = 19, 
S = 50$  $S = 100$ subjects, respectively.

The two figures show that the PMwG and \dt_SMC{} estimators give very similar results---suggesting that the two algorithms are correctly implemented. The posterior densities narrow as the number of subjects $S$ increases, and the posterior modes of the parameters 
are consistent with the true data generating values.
Each panel in the two figures also includes a third posterior distribution (in yellow), which illustrates
the effect of making 
the standard hierarchical LBA assumption that the random effects are independent, using the same priors
for the group level parameters as in the correlated case discussed above, and using PMwG to sample.%\footnote{read through the immediate above. I don't think the independent prior we now use is the same as that used before. 
%\cite{Turner2013} has a truncated normal on the untransformed random effects; please correct if I am wrong.}

%normal distributions for the random effects, since \cite{Turner2013}. We used the same priors for the group level parameters %for this uncorrelated fit as for the others, and used PMwG to draw samples. 

The results for this simplified, uncorrelated, LBA model differ systematically to the LBA model proposed here. The posterior mean estimates are more certain, 
that is, the distributions are more peaked (Figure \ref{fig:kerneldensitysimthetaS19T1000}), and the posterior variance
estimates are smaller (Figure \ref{fig:kerneldensitysimsigmaS19T1000}). This suggests that using a model with
independent distributions for the random effects, as is standard in the literature, can -- if there really is correlation in the data -- lead to unwarranted overconfidence
in estimation precision, and under-estimation of the magnitude of individual differences.

We estimated the marginal likelihood for each simulated experiment ($S=19$, $S=50$, and $S=100$ subjects)
using the outputs of the \dt_SMC{} sampler and  the three estimators: the standard \dt_SMC{} method, and the 
thermodynamic integration estimators $TI_1$ and $TI_2$ discussed in 
Section~\ref{MarginalLikelihoodEstimation}. 
Table~\ref{MarginalLikelihoodTable} reports the logs of the marginal likelihood estimates (with the standard error in
brackets); the standard errors are obtained using ten replicates for each estimator. The three methods agree very closely, and the standard errors for each method are very small, 
suggesting that the log of the marginal likelihoods are estimated accurately.
%\footnote{DG: I put in this sentence. Do you agree with it? Is it redunadant?}

%compared with the differences in the estimates of the marginal likelihoods between different models that are usually considered scientifically meaningful

\begin{table}[h]
\caption{Logs of the marginal likelihood estimates (with standard errors in
brackets). \label{MarginalLikelihoodTable}}

\begin{centering}
\begin{tabular}{ccccc}
\hline 
S & N & DT-SMC & $TI_{1}$ & $TI_{2}$\tabularnewline
\hline 
19 & 1,000 & $\underset{\left(2.33\right)}{8,219.29}$ & $\underset{\left(2.21\right)}{8,219.05}$ & $\underset{\left(2.21\right)}{8,219.53}$\tabularnewline
50 & 1,000 & $\underset{\left(7.72\right)}{25,837.82}$ & $\underset{\left(7.79\right)}{25,837.70}$ & $\underset{\left(7.79\right)}{25,838.19}$\tabularnewline
100 & 1,000 & $\underset{\left(13.03\right)}{44,370.31}$ & $\underset{\left(13.00\right)}{44,370.68}$ & $\underset{\left(12.99\right)}{44,371.19}$\tabularnewline
\hline 
\end{tabular}
\par\end{centering}
\end{table}

\begin{figure}[h]
\centering{}\includegraphics[width=13cm,height=6cm]{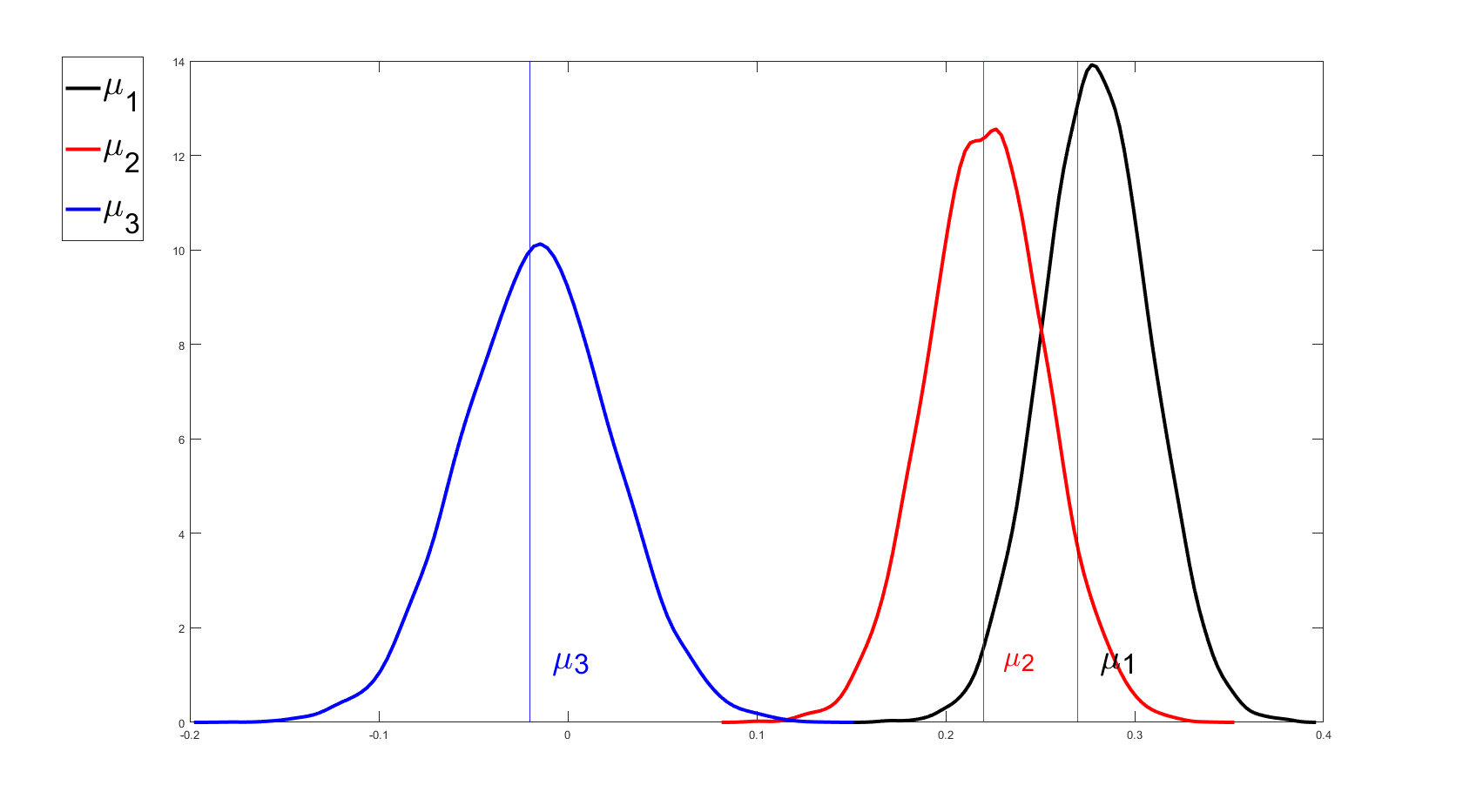}
\caption{Kernel density estimates of the marginal posterior densities over the decision threshold parameters
$\muab{\mu}{1}{}$, $\muab{\mu}{2}{}$
and $\muab{\mu}{3}{}$ obtained using the PMwG sampler in the simulation study with $S=100$ subjects 
and $N=1,000$.
The vertical lines show the true (data-generating) values. \label{fig:Kernel-density-estimates marginal posterior density of mean threshold parameters} }
\end{figure}

Figure~\ref{fig:Kernel-density-estimates marginal posterior density of mean threshold parameters} 
shows the kernel density estimates of the marginal 
posterior densities for the three parameters $\muab{\mu}{1}{}$,
$\muab{\mu}{2}{}$ and $\muab{\mu}{3}{}$ 
which govern the decision threshold in the accuracy, neutral, and speed emphasis conditions (respectively)
of the simulated experiment. There is considerable overlap between the two marginal distributions representing the different threshold parameters for the accuracy and neutral conditions. If these data were from a real (not simulated) experiment, this might be interpreted as evidence that the participants in the experiment failed to distinguish between those two conditions; that they did (or could) not adopt different decision-making styles when asked to.
%\footnote{SB and GH:
%\begin{itemize}
%   \item 
%This is a somewhat confusing discussion because we KNOW that the data is generated with three different conditions. So if we had enough data we would likely find the differences. You are arguing here like a frequentist would argue that the null hypothesis is that there are two conditions and we would not reject the null unless there was strong evidence against it.
%\item 
%I find this to be a somewhat fallacious argument, as it is just ``data mining". A better approach may be to do a simulated example with just two conditions and show that model selection via marginal likelihood works. 
%\end{itemize}
%FROM SB: I have revised some of the offending text. I can see the problem. I've tried to use this as an example of how one %might use the results of the new methods for applied investigation.
%}

This is exactly the kind of model selection question facing researchers using the LBA model. \secref{subsec:Real-Applications} demonstrates how to use the marginal likelihood estimates obtained using
\dt_SMC{} to further investigate whether participants adopted different decision-making thresholds in the three conditions. In that section, we compare the unrestricted (three parameter) model against restricted models having 
shared threshold parameters. An alternative approach to this question could enforce order constraints 
on the estimated random effects. This is easily accomplished by parameter transformations -- define the speed-emphasis threshold as usual, and then estimate parameters for the differences between that threshold and the neutral condition threshold, and between the neutral and accuracy condition thresholds. These incremental parameters can be constrained to be positive by estimating their logarithms, as with the other parameters, and inferential tests made by comparing the estimated increments with zero. 

\subsection*{Comparing the PMwG sampler to existing estimation methods}

The hierarchical LBA model is most often estimated using DE-MCMC as developed by \citet{Turner2013}. 
However, such applications have been restricted to an LBA model which assumes independent truncated 
normal prior distributions for the random effects, whereas the LBA model defined in 
Section~\ref{sec:The-Linear-Ballistic accumulator} allows for a (correlated) multivariate 
normal prior distribution on the log of the random effects. The considerably simpler model estimated via DE-MCMC
does not require the estimation of covariance elements. 
An uncorrelated model with $D_\alpha$ random effects per person 
has $2D_\alpha$ group-level parameters 
(a mean and a standard deviation for each random effect). In contrast, the current LBA model 
has $D_\alpha+(D_\alpha+1)D_\alpha/2$ parameters (a mean and standard deviation for each random effect, 
plus all pairwise covariances).  \ref{DetailComparisonPMwG} shows that the PMwG sampler agrees with the DE-MCMC sampler, when estimating the simplified LBA model without correlations.

Estimating the LBA model with non-zero correlations in the prior may be impossible with the current DE-MCMC
sampler and will require extensive modifications at a minimum. The DE-MCMC sampler generates the group level mean and standard deviation for each of the random effects blocked in pairs. This blocking breaks down when the off-diagonal covariances in the prior are non-zero, 
as they are highly correlated with the (diagonal) variance parameters, and so some other blocking scheme will need to be devised. It is not clear what that scheme should be. Further, DE-MCMC generates proposals by taking linear combinations of existing samples. When the correlations are non-zero, this means generating covariance matrix proposals, but the differential evolution approach does not ensure that the proposals will be valid covariance matrices (positive definite). It is also 
standard to set the tuning parameter in the DE-MCMC algorithm inversely proportional to
the number of parameters. This is likely to make the method very inefficient given the much larger number 
of unknown parameters when a full covariance matrix is used in the prior (scales with $D_{\alpha}^2$).
It is possible that these difficulties for DE-MCMC may be overcome with its further development. However, such an extension of the DE-MCMC approach is beyond the scope of this paper.

In contrast, PMwG scales well as the number of parameters increases because it uses Gibbs sampling to generate the group-level parameters. In demonstrating this scaling property, \citet{Gunawan:2019b} 
extend the PMwG approach to estimate a high-dimensional hierarchical LBA model with $S=110$ subjects, $N=1,350$ trials, 
and $D_{\alpha}=30$ individual random effects parameters---giving  a covariance matrix with 435 free parameters. 
\citet{Gunawan:2019a} extend the hierarchical LBA model to allow the individual level parameters of 
each subject to change over blocks of the trials,
thus extending the PMwG approach to estimate time-varying LBA models. 
Both these extensions show that the methods proposed in this article
allow exploration of important psychological questions that are at present neglected due to computational  intractability. 

\newcommand{\mualnb}[3]{{\bs {#1}}_{LN,\alpha_{{#2}^{#3}}}}
\newcommand{\sigmalnb}[1]{{\bs \Sigma}_{LN,\alpha,{#1}}}

\subsection{Real Data Application}\label{subsec:Real-Applications}

We applied PMwG and \dt_SMC{} to the behavioural data first presented by \citet{Forstmann2008}, and introduced in Section \ref{sec:The-Linear-Ballistic accumulator}. PMwG and \dt_SMC{} were run with the same settings as in \ref{tuningPMwG} and \ref{tuningAISIL}, respectively. The wall-clock computation times to run PMwG and \dt_SMC{}
were around 30 and 138 minutes, respectively, using a Matlab implementation of the algorithm and 28 CPU-cores.

Table~\ref{tab:Inefficiency-Factors-of LBA parameters real1} shows the estimated group-level parameters 
from both methods, along with the estimated posterior standard deviations and the IACT inefficiency factors for 
PMwG. All the IACT values are again small, indicating that the chains mixed well and that the performance 
of the sampler did not deteriorate markedly when moving from simulated data to real data. 
The  posterior mean estimates and the associated posterior standard deviations from PMwG and \dt_SMC{} 
are very close to each other, providing an accuracy check on both samplers.

\begin{table}[h] 
\caption{Posterior means (with posterior standard deviations in brackets) and IACT of the LBA group-level parameters of the full model estimated
using PMwG and \dt_SMC{} (DT-SMC) for the data from Forstmann et al. (2008). The order of the random effect parameters in the covariance matrix $\bs \Sigma$ is ${b^{(1)}}$,
${b^{(2)}}$, ${b^{(3)}}$, ${A}$,
${v^{(1)}}$, ${v^{(2)}}$, and
${\tau}$. 
\label{tab:Inefficiency-Factors-of LBA parameters real1}}
{\tiny
\centering{}{\small{}}%
\begin{tabular}{cccccccc}
\tabularnewline\hline\tabularnewline
{\small{}Param.} & {\small{}$\textrm{Est}$} & {\small{}$\textrm{IACT}$} & {\small{}$\textrm{Est}$} & {\small{}Param.} & {\small{}$\textrm{Est}$} & {\small{}$\textrm{IACT}$} & {\small{}$\textrm{Est}$}\tabularnewline
\hline\tabularnewline
 & {\small{}PMwG} & \small{}PMwG & {\small{}DT-SMC} &  & {\small{}PMwG} & \small{}PMwG & \small{}DT-SMC \tabularnewline
\hline\tabularnewline
{\small{}$\muab{\mu}{1}{}$} & {\small{}$\underset{\left(0.06\right)}{0.27}$} & {\small{}$1.22$} & {\small{}$\underset{\left(0.06\right)}{0.28}$} & {\small{}$\sigmab{11}$} & {\small{}$\underset{\left(0.03\right)}{0.06}$} & {\small{}$1.68$} & {\small{}$\underset{\left(0.03\right)}{0.06}$}\tabularnewline
{\small{}$\muab{\mu}{2}{}$} & {\small{}$\underset{\left(0.06\right)}{0.22}$} & {\small{}$1.17$} & {\small{}$\underset{\left(0.06\right)}{0.22}$} & {\small{}$\sigmab{22}$} & {\small{}$\underset{\left(0.03\right)}{0.07}$} & {\small{}$1.62$} & {\small{}$\underset{\left(0.03\right)}{0.07}$}\tabularnewline
{\small{}$\muab{\mu}{3}{}$} & {\small{}$\underset{\left(0.08\right)}{-0.02}$} & {\small{}$1.10$} & {\small{}$\underset{\left(0.09\right)}{-0.01}$} & {\small{}$\sigmab{33}$} & {\small{}$\underset{\left(0.05\right)}{0.13}$} & {\small{}$1.49$} & {\small{}$\underset{\left(0.05\right)}{0.13}$}\tabularnewline
{\small{}$\muab{\mu}{4}{}$} & {\small{}$\underset{\left(0.07\right)}{-0.40}$} & {\small{}$1.49$} & {\small{}$\underset{\left(0.07\right)}{-0.40}$} & {\small{}$\sigmab{44}$} & {\small{}$\underset{\left(0.04\right)}{0.09}$} & {\small{}$2.61$} & {\small{}$\underset{\left(0.04\right)}{0.09}$}\tabularnewline
{\small{}$\muab{\mu}{5}{}$} & {\small{}$\underset{\left(0.11\right)}{0.30}$} & {\small{}$1.26$} & {\small{}$\underset{\left(0.11\right)}{0.31}$} & {\small{}$\sigmab{55}$} & {\small{}$\underset{\left(0.09\right)}{0.22}$} & {\small{}$2.00$} & {\small{}$\underset{\left(0.08\right)}{0.22}$}\tabularnewline
{\small{}$\muab{\mu}{6}{}$} & {\small{}$\underset{\left(0.04\right)}{1.12}$} & {\small{}$1.28$} & {\small{}$\underset{\left(0.04\right)}{1.13}$} & {\small{}$\sigmab{66}$} & {\small{}$\underset{\left(0.02\right)}{0.03}$} & {\small{}$2.53$} & {\small{}$\underset{\left(0.02\right)}{0.03}$}\tabularnewline
{\small{}$\muab{\mu}{7}{}$} & {\small{}$\underset{\left(0.07\right)}{-1.74}$} & {\small{}$2.16$} & {\small{}$\underset{\left(0.07\right)}{-1.75}$} & {\small{}$\sigmab{77}$} & {\small{}$\underset{\left(0.04\right)}{0.09}$} & {\small{}$5.70$} & {\small{}$\underset{\left(0.04\right)}{0.09}$}\tabularnewline\hline\tabularnewline
{\footnotesize{}$\sigmab{12}$} & {\footnotesize{}$\underset{\left(0.03\right)}{0.06}$} & {\footnotesize{}$1.66$} & {\footnotesize{}$\underset{\left(0.03\right)}{0.07}$} & {\footnotesize{}$\sigmab{34}$} & {\footnotesize{}$\underset{\left(0.04\right)}{0.08}$} & {\footnotesize{}$1.53$} & {\footnotesize{}$\underset{\left(0.04\right)}{0.08}$} \tabularnewline
{\footnotesize{}$\sigmab{13}$} & {\footnotesize{}$\underset{\left(0.03\right)}{0.08}$} & {\footnotesize{}$1.61$} & {\footnotesize{}$\underset{\left(0.04\right)}{0.08}$} & {\footnotesize{}$\sigmab{35}$} & {\footnotesize{}$\underset{\left(0.05\right)}{0.11}$} & {\footnotesize{}$1.59$} & {\footnotesize{}$\underset{\left(0.05\right)}{0.11}$} \tabularnewline
{\footnotesize{}$\sigmab{14}$} & {\footnotesize{}$\underset{\left(0.03\right)}{0.06}$} & {\footnotesize{}$1.63$} & {\footnotesize{}$\underset{\left(0.03\right)}{0.06}$} & {\footnotesize{}$\sigmab{36}$} & {\footnotesize{}$\underset{\left(0.02\right)}{0.01}$} & {\footnotesize{}$1.56$} & {\footnotesize{}$\underset{\left(0.02\right)}{0.01}$} \tabularnewline
{\footnotesize{}$\sigmab{15}$} & {\footnotesize{}$\underset{\left(0.04\right)}{0.06}$} & {\footnotesize{}$1.70$} & {\footnotesize{}$\underset{\left(0.04\right)}{0.07}$} & {\footnotesize{}$\sigmab{37}$} & {\footnotesize{}$\underset{\left(0.04\right)}{-0.09}$} & {\footnotesize{}$2.89$} & {\footnotesize{}$\underset{\left(0.04\right)}{-0.09}$} \tabularnewline
{\footnotesize{}$\sigmab{16}$} & {\footnotesize{}$\underset{\left(0.01\right)}{0.00}$} & {\footnotesize{}$1.61$} & {\footnotesize{}$\underset{\left(0.01\right)}{0.00}$} & {\footnotesize{}$\sigmab{45}$} & {\footnotesize{}$\underset{\left(0.04\right)}{0.04}$} & {\footnotesize{}$1.70$} & {\footnotesize{}$\underset{\left(0.04\right)}{0.04}$} \tabularnewline
{\footnotesize{}$\sigmab{17}$} & {\footnotesize{}$\underset{\left(0.03\right)}{-0.05}$} & {\footnotesize{}$2.67$} & {\footnotesize{}$\underset{\left(0.03\right)}{-0.05}$} & {\footnotesize{}$\sigmab{46}$} & {\footnotesize{}$\underset{\left(0.01\right)}{0.00}$} & {\footnotesize{}$1.60$} & {\footnotesize{}$\underset{\left(0.02\right)}{0.00}$} \tabularnewline
{\footnotesize{}$\sigmab{23}$} & {\footnotesize{}$\underset{\left(0.04\right)}{0.09}$} & {\footnotesize{}$1.56$} & {\footnotesize{}$\underset{\left(0.04\right)}{0.09}$} & {\footnotesize{}$\sigmab{47}$} & {\footnotesize{}$\underset{\left(0.03\right)}{-0.05}$} & {\footnotesize{}$2.00$} & {\footnotesize{}$\underset{\left(0.03\right)}{-0.05}$} \tabularnewline
{\footnotesize{}$\sigmab{24}$} & {\footnotesize{}$\underset{\left(0.03\right)}{0.06}$} & {\footnotesize{}$1.56$} & {\footnotesize{}$\underset{\left(0.03\right)}{0.06}$} & {\footnotesize{}$\sigmab{56}$} & {\footnotesize{}$\underset{\left(0.02\right)}{0.01}$} & {\footnotesize{}$1.58$} & {\footnotesize{}$\underset{\left(0.02\right)}{0.01}$} \tabularnewline
{\footnotesize{}$\sigmab{25}$} & {\footnotesize{}$\underset{\left(0.04\right)}{0.08}$} & {\footnotesize{}$1.72$} & {\footnotesize{}$\underset{\left(0.04\right)}{0.08}$} & {\footnotesize{}$\sigmab{57}$} & {\footnotesize{}$\underset{\left(0.05\right)}{-0.09}$} & {\footnotesize{}$2.82$} & {\footnotesize{}$\underset{\left(0.05\right)}{-0.10}$} \tabularnewline
{\footnotesize{}$\sigmab{26}$} & {\footnotesize{}$\underset{\left(0.01\right)}{0.01}$} & {\footnotesize{}$1.71$} & {\footnotesize{}$\underset{\left(0.01\right)}{0.01}$} & {\footnotesize{}$\sigmab{67}$} & {\footnotesize{}$\underset{\left(0.01\right)}{0.00}$} & {\footnotesize{}$1.57$} & {\footnotesize{}$\underset{\left(0.01\right)}{0.00}$} \tabularnewline
{\footnotesize{}$\sigmab{27}$} & {\footnotesize{}$\underset{\left(0.03\right)}{-0.06}$} & {\footnotesize{}$3.05$} & {\footnotesize{}$\underset{\left(0.03\right)}{-0.06}$} & & & \tabularnewline\hline
\end{tabular}{\small \par}
}
\end{table}

\begin{figure}[h] 
\centering{}\includegraphics[width=13cm,height=10cm]{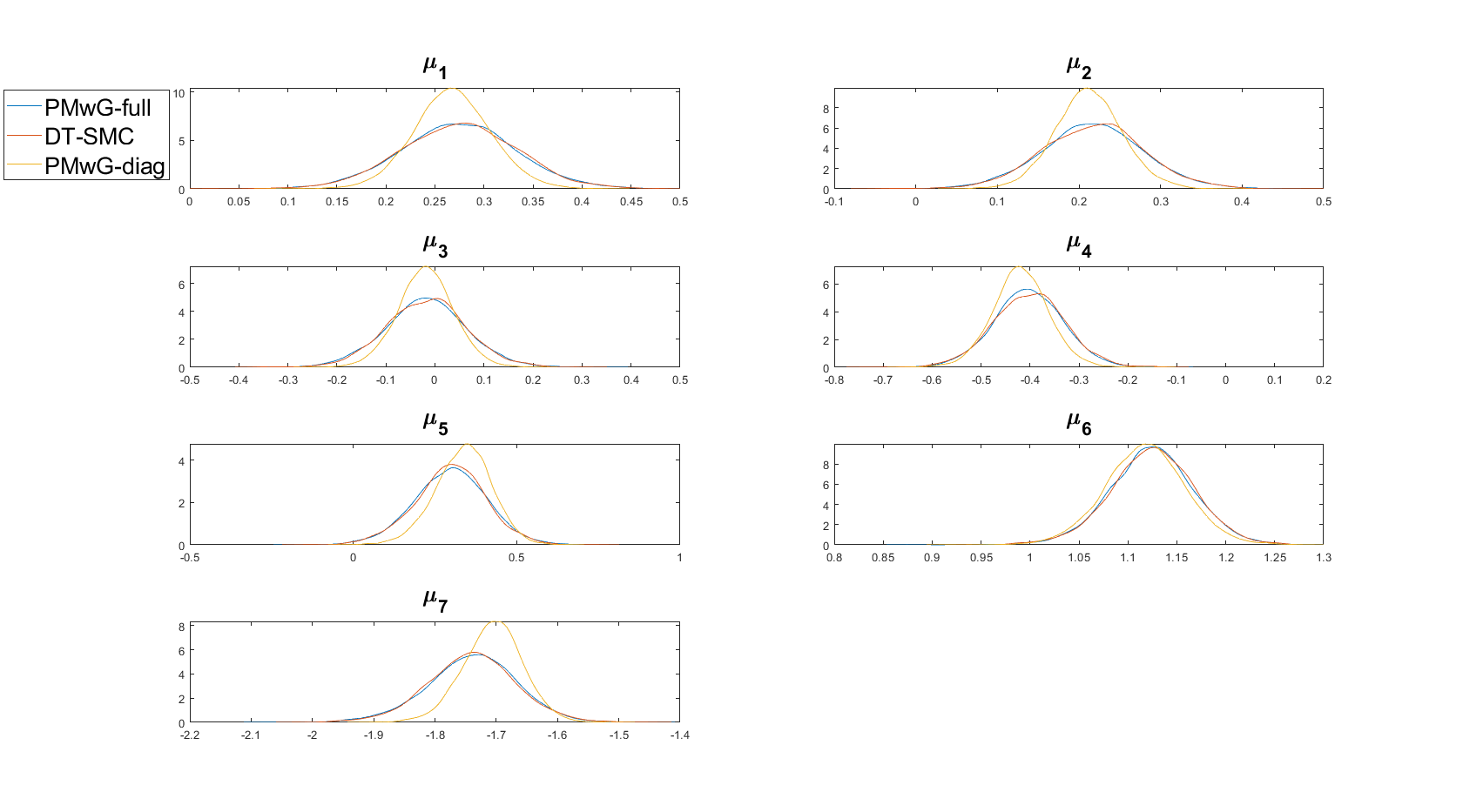}
\caption{Kernel density estimates of the LBA group-level mean parameters ($\mu$) for the data presented by \citet{Forstmann2008}.
\label{fig:Kernel-Density-Estimate of LBA real parameters mu}}
\end{figure}

\begin{figure}[h] 
\centering{}\includegraphics[width=13cm,height=10cm]{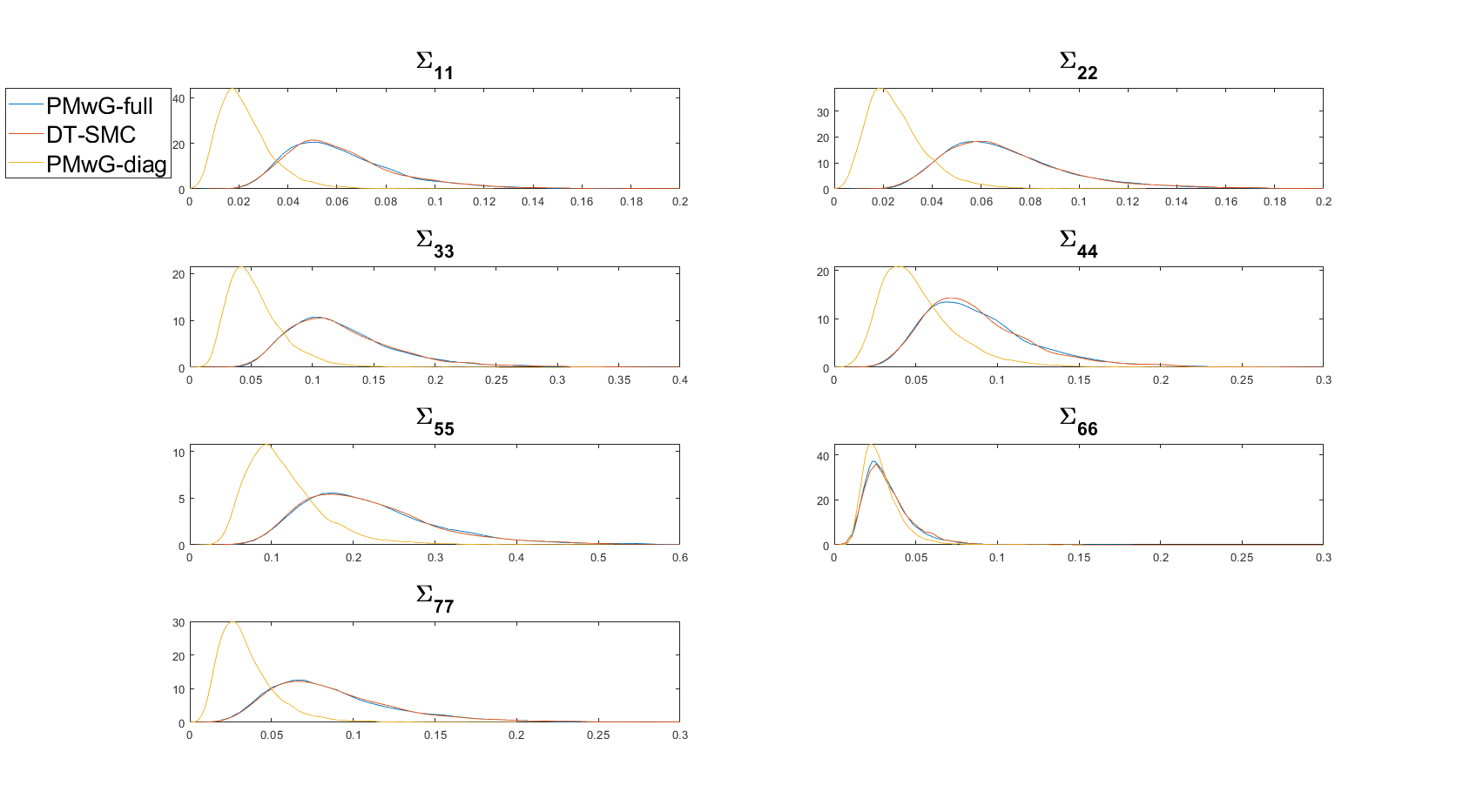}
\caption{Kernel density estimates of the LBA group-level variance parameters (diagonal elements of $\Sigma$) for the \citet{Forstmann2008} data.
\label{fig:Kernel-Density-Estimate of LBA real parameters sigma}}
\end{figure}

Figure~\ref{fig:Kernel-Density-Estimate of LBA real parameters mu} shows the 
kernel density estimates of marginal posterior densities for the group mean parameters, using samples from PMwG (blue) and
\dt_SMC{} (red); the figure also shows the estimated posterior distributions for the reduced model, 
with zero prior correlation between the individual level parameters estimated using PMwG (yellow). 
Figure~\ref{fig:Kernel-Density-Estimate of LBA real parameters sigma} shows the corresponding results
for the between-subject variance parameters instead of the means, 
i.e., the diagonal elements of $\Sigma$. 
Both figures demonstrate that the \dt_SMC{} estimates are very close to the PMwG estimates, for all the parameters 
for the LBA model with a full prior covariance matrix for the 
random effects. However, similarly to the simulation study, the posterior estimates 
from the simplified LBA model (with zero covariance) are quite different to the posterior estimates 
from the LBA model with a full covariance matrix. The differences are particularly pronounced for the variance parameters in
Figure~\ref{fig:Kernel-Density-Estimate of LBA real parameters sigma}. In all cases, the zero-covariance reduced model
estimates are much smaller for the between-subjects variances, and often also have narrower posterior distributions. 
This matches the results from the simulation study above, and is consistent with the hypothesis that the real data includes
non-zero between-subject correlations, and failing to take this into account in the prior leads 
to unwarranted overconfidence in the posteriors.

%Figures \ref{fig:kerneldensitysimthetaS19T1000} and \ref{fig:kerneldensitysimsigmaS19T1000} show the kernel density estimates of some of the LBA group-level parameters, the mean $\mu_{\alpha}$ and the diagonal elements of the covariance matrix $\Sigma_{\alpha}$, based on PMwG and DTSMC-RE samples for $S=19$ subjects for the hierarchical LBA model defined in Section \ref{subsec:Hierarchical-Bayesian-Model}. The figure shows that the DTSMC-RE estimates are very close to the PMwG estimates for all parameters, which provides a check on the consistency of the two algorithms. The figure also shows the kernel density estimates of the mean $\mu_{\alpha}$ and the diagonal elements of the covariance matrix $\Sigma_{\alpha}$ assuming independent normal distributions for the individual random effects as in \citet{Turner2013}. Same priors are used in order to make valid comparisons between two models. The posterior estimates from the simplified LBA model are quite different to the posterior estimates from the LBA model with full covariance matrix. The two LBA models have similar posterior means for the group level means $\mu_{\alpha}$ for all random effects, but the posterior variances of the simplified LBA model is smaller. It is also clear from Figure \ref{fig:kerneldensitysimsigmaS19T1000} that the posterior estimates from the simplified LBA model have much smaller posterior means and variances for the diagonal elements of the covariance matrix $\Sigma_{\alpha}$.

%\begin{figure}[h]
%\caption{Kernel density estimates some of the individual level random effects
%parameters of the full model \label{fig:Kernel-Density-Estimate of LBA real random effects}}
%\centering{}\includegraphics[width=17cm,height=20cm]{ksdensityPMCMCAISILlatent}
%\end{figure}

The model setup in Section \ref{sec:The-Linear-Ballistic accumulator} allows us to  estimate
the correlation matrix $\Gamma$ between individual level parameters by using the standard transformation of the estimated covariance matrix.
%\begin{align} \label{eq: correln}
%\Gamma &=\textrm{diag}\left(\bs \Sigma_{LN,\alpha}\right)^{-\frac{1}{2}}\bs \Sigma_{LN,\alpha}\textrm{diag}\left(\bs
%\Sigma_{LN,\alpha}\right)^{-\frac{1}{2}}.
%\end{align}
Table~\ref{tab:pos mean correlation parameters full model} shows that the threshold parameters for all three conditions are highly correlated: $\Gamma\left(b^{(1)},b^{(2)}\right)=.96$, $\Gamma\left(b^{(1)},b^{(3)}\right)=.87$, and $\Gamma\left(b^{(2)},b^{(3)}\right)=.93$. The maximum value of the start point distribution ($A$) is also highly correlated with the threshold parameters; $\Gamma\left(b^{(1)},A\right)=.80$, $\Gamma\left(b^{(2)},A\right)=.76$, and $\Gamma\left(b^{(3)},A\right)=.70$. The non-decision time parameters at the individual subject level are negatively correlated with all other individual level parameters -- presumably because of trade-offs between explaining the same RT as either composed of more or less decision time vs. non-decision time. The mean drift rate for the accumulator corresponding to the correct response is not highly correlated with other individual level parameters. The magnitudes of these correlations emphasize the importance of explicitly modelling the covariance matrix, rather than forcing it to have 
zero correlations.
%\footnote{Table 5 is Ok as we are looking for interpretation, but I would write $\Gamma(b^{(1)}, b^{2}) $ etc 
%as you wrote in text above to  make the text and the table consistent. I tried on first row and it looks OK. Instead of the heading pair please put Param.}

\begin{table}[h]
\caption{Posterior means (with posterior standard deviations in brackets) of the correlations
of the LBA parameters obtained using \dt_SMC{}. \label{tab:pos mean correlation parameters full model}}
{\small
\centering{}%
\begin{tabular}{cccccc}
\hline
Param & Est. & Param & Est. & Param & Est.\tabularnewline
\hline
$\Gamma{\left(b^{\left(1\right)},b^{\left(2\right)}\right)}$ & $\underset{\left(0.02\right)}{0.96}$ & $ \Gamma{\left(b^{\left(2\right)},A\right)}$ & $\underset{\left(0.12\right)}{0.76}$ & $\Gamma {\left(b^{\left(3\right)},\tau\right)}$ & $\underset{\left(0.06\right)}{-0.70}$\tabularnewline
$\Gamma{\left(b^{\left(1\right)},b^{\left(3\right)}\right)}$ & $\underset{\left(0.05\right)}{0.87}$ & $ \Gamma{\left(b^{\left(2\right)},v^{\left(1\right)}\right)}$ & $\underset{\left(0.14\right)}{0.57}$ & $ \Gamma{\left(A,v^{\left(1\right)}\right)}$ & $\underset{\left(0.20\right)}{0.27}$\tabularnewline
$ \Gamma{\left(b^{\left(1\right)},A\right)}$ & $\underset{\left(0.10\right)}{0.80}$ &  $\Gamma {\left(b^{\left(2\right)},v^{\left(2\right)}\right)}$ & $\underset{\left(0.23\right)}{0.09}$ & $\Gamma {\left(A,v^{\left(2\right)}\right)}$ & $\underset{\left(0.23\right)}{-0.03}$\tabularnewline
$\Gamma {\left(b^{\left(1\right)},v^{\left(1\right)}\right)}$ & $\underset{\left(0.16\right)}{0.49}$ & $ \Gamma{\left(b^{\left(2\right)},\tau\right)}$ & $\underset{\left(0.07\right)}{-0.69}$ & $ \Gamma{\left(A,\tau\right)}$ & $\underset{\left(0.16\right)}{-0.46}$\tabularnewline
$\Gamma {\left(b^{\left(1\right)},v^{\left(2\right)}\right)}$ & $\underset{\left(0.23\right)}{0.03}$ & $\Gamma {\left(b^{\left(3\right)},A\right)}$ & $\underset{\left(0.14\right)}{0.70}$ & $ \Gamma{\left(v^{\left(1\right)},v^{\left(2\right)}\right)}$ & $\underset{\left(0.22\right)}{0.16}$\tabularnewline
$\Gamma {\left(b^{\left(1\right)},\tau\right)}$ & $\underset{\left(0.09\right)}{-0.63}$ & $ \Gamma{\left(b^{\left(3\right)},v^{\left(1\right)}\right)}$ & $\underset{\left(0.14\right)}{0.62}$ & $ \Gamma{\left(v^{\left(1\right)},\tau\right)}$ & $\underset{\left(0.09\right)}{-0.55}$\tabularnewline
$\Gamma {\left(b^{\left(2\right)},b^{\left(3\right)}\right)}$ & $\underset{\left(0.03\right)}{0.93}$ & $ \Gamma{\left(b^{\left(3\right)},v^{\left(2\right)}\right)}$ & $\underset{\left(0.23\right)}{0.14}$ & $ \Gamma{\left(v^{\left(2\right)},\tau\right)}$ & $\underset{\left(0.22\right)}{-0.03}$\tabularnewline
\hline
\end{tabular}
}
\end{table}

Table~\ref{tab:pos mean LBA parameters full model} summarizes the posterior mean estimates of the parameters on the original
(not logarithmic) scale obtained using \eqref{eq: log normal distn}. These values suggest that there may only be 
slight differences in the estimates of the threshold parameters between the three conditions: accuracy, neutral, and speed ($\tilde{{\mu}}_{1}$, $\tilde{{\mu}}_{2}$, and $\tilde{{\mu}}_{3}$). To investigate this, we estimated a
restricted model with two threshold parameters by combining the accuracy and neutral conditions, as well as
a more restricted model with a single, shared, threshold parameter for all three conditions. We used
\dt_SMC{},  specified as above, to estimate the marginal likelihood for each model.
%Estimation results of the restricted LBA models and detailed explanations can be found in the supplement \citep{Gunawan2018}. 
Table~\ref{marglikreal} reports the estimated log marginal likelihoods (with standard errors in brackets) 
for the three models. The differences between the log marginal likelihoods are much larger than the standard errors, and also large
relative to the scales usually used to judge statistical reliability, e.g.
the corresponding Bayes factors for the model comparisons are all much larger than $10^6$. 
The results favour the unrestricted model for these data, which also supports the analyses by 
\citet{Forstmann2008}. An important caveat for this analysis is that the marginal likelihoods -- for any model -- depend on the prior. For example, another approach to the problem would set up the random effects using an intercept plus effect coding, and specify a prior distribution for the effect sizes. Further work will be required to investigate the sensitivity of model selection outcomes to different choices of prior distribution.

\begin{table}[H]
\caption{Posterior means (with posterior standard deviations in brackets) of the group level
LBA parameters of the model with three threshold
parameters on the original
(not logarithmic) scale obtained using \eqref{eq: log normal distn}. The order of the 
random effect parameters is the same as defined in Section \ref{sec:The-Linear-Ballistic accumulator}.
\label{tab:pos mean LBA parameters full model}}
{
\centering{}
\begin{center}\begin{tabular}{cccc}\tabularnewline
\hline
Param. & Est. & Param & Est. \tabularnewline
\hline
$\tilde{{\mu}}_{1}$ & $\underset{\left(0.08\right)}{1.36}$ & $\tilde{\Sigma}_{1,1}$ & $\underset{\left(0.06\right)}{0.13}$ \tabularnewline
$\tilde{{\mu}}_{2}$ & $\underset{\left(0.09\right)}{1.30}$ & $\tilde{\Sigma}_{2,2}$ & $\underset{\left(0.06\right)}{0.13}$ \tabularnewline
$\tilde{{\mu}}_{3}$ & $\underset{\left(0.09\right)}{1.06}$ & $\tilde{\Sigma}_{3,3}$ & $\underset{\left(0.09\right)}{0.16}$ \tabularnewline
$\tilde{{\mu}}_{4}$ & $\underset{\left(0.05\right)}{0.70}$ & $\tilde{\Sigma}_{4,4}$ & $\underset{\left(0.03\right)}{0.05}$ \tabularnewline
$\tilde{{\mu}}_{5}$ & $\underset{\left(0.18\right)}{1.52}$ & $\tilde{\Sigma}_{5,5}$ & $\underset{\left(0.37\right)}{0.60}$ \tabularnewline
$\tilde{{\mu}}_{6}$ & $\underset{\left(0.14\right)}{3.14}$ & $\tilde{\Sigma}_{6,6}$ & $\underset{\left(0.17\right)}{0.34}$ \tabularnewline
$\tilde{{\mu}}_{7}$ & $\underset{\left(0.01\right)}{0.18}$ & $\tilde{\Sigma}_{7,7}$ & $\underset{\left(0.002\right)}{0.003}$ \tabularnewline
\hline
\end{tabular}
\end{center}
}
\end{table}

\begin{table}[H]
\caption{Log of the marginal likelihood estimates (with standard errors in
brackets), for the three models with different numbers of free threshold parameters (1, 2, or 3).\label{marglikreal}}

\centering{}%
\begin{tabular}{cccc}
\hline 
Thresholds & Standard \dt_SMC{} & $TI_{1}$ & $TI_{2}$\tabularnewline
\hline 
One & $\underset{\left(1.59\right)}{5,200.58}$ & $\underset{\left(1.52\right)}{5,200.48}$ & $\underset{\left(1.52\right)}{5,200.99}$\tabularnewline
Two & $\underset{\left(2.38\right)}{7,350.80}$ & $\underset{\left(2.24\right)}{7,350.56}$ & $\underset{\left(2.23\right)}{7,351.05}$\tabularnewline
Three & $\underset{\left(3.10\right)}{7,447.31}$ & $\underset{\left(3.04\right)}{7,447.28}$ & $\underset{\left(3.04\right)}{7,447.75}$\tabularnewline
\hline 
\end{tabular}
\end{table}

\section{Conclusions\label{sec:Conclusions}}

Based on recent advances in particle MCMC,
the article develops two new estimation approaches for the Linear
Ballistic Accumulator model of \citet{Brown2008}; 
Particle Metropolis within Gibbs and density tempered SMC.
We show that PMwG and \dt_SMC{} perform well for both simulated and real data.  The new methods are alternatives to the existing approach that is based on MCMC with proposals generated by differential evolution 
\citep{Turner2013} and provide important advantages. 
%PMwG produces iterates with much smaller autocorrelation than DE-MCMC, and hence increased statistical sampling efficiency.
Density tempered SMC is extremely well-suited to parallelisation on high-performance computers, which is likely to be an advantage in future work with large-scale models and data. Although both PMwG and DE-MCMC are also both appropriate for parallelisation in high-performance computing environments, DE-MCMC requires more frequent dependence between the multiple chains, which limits its efficiency when parallelised. 
We also found that the PMwG sampler is much faster than \dt_SMC{} and can be easily implemented with minimal computational resources, e.g., personal computers. % However, \dt_SMC{} is easier to parallelise than PMwG, so that it is likely to be faster than PMwG when very powerful computing resources are available and/or the analysis problem is large. 
Furthermore, \dt_SMC{} provides an estimate of the marginal likelihood at little extra cost, 
and hence it can be readily used for model selection via Bayes factors.

Another important contribution of our work is to explicitly model the full covariance structure of the prior for 
the random effects. Like all plausible cognitive models, there are substantial correlations between the individual level parameters of the LBA model: subjects with a large decision threshold also tend to have a large starting point distribution, etc. In previous applications of the model, 
these prior correlations were set to zero, 
with the group-level distributions treated \emph{a priori} as independent. Despite making this \emph{a priori} assumption of independence, the resulting posterior samples always exhibited strong posterior correlations between individual level parameters. Explicitly allowing non-zero correlations in the prior, as we have done, provides better estimates of the parameters and their variances, and improves computational efficiency.

The computational flexibility of 
the new methods allows exploration of important psychological questions which have hitherto been neglected, 
due to statistical intractability. 
For example, it is well known that there can be substantial sequential effects in decision-making data: both response choices and response times tend to be positively autocorrelated. All applications of the LBA model -- and
indeed, almost all decision-making models -- have ignored these sequential effects, treating the data as i.i.d. and attributing the effects of any sequential dependence to error terms. Both our approaches allow tractable extensions that explicitly take into 
account of within subject dependence and other interesting sequential effects such as
parameter evolution due to fatigue or learning. We are investigating these models in ongoing work.

To aid researchers in adopting the proposed methods, we provide scripts that implement
both the PMwG and \dt_SMC{} methods as applied to the real data from \citet{Forstmann2008}; 
see \url{osf.io/5b4w3} for more details.

\section*{Acknowledgements}

We thank three anonymous referees and an Action editor for comments 
that improved the scientific content and clarity of the paper. 
The research of Gunawan, Tran, Kohn and Brown was partially 
supported by the Australian Research Council (ARC) Discovery grant DP180102195; 
Gunawan and Kohn were also supported by ARC Discovery grant DP150104630; and Hawkins by the ARC DECRA grant DE170100177.

\section{References}
\bibliographystyle{elsarticle-harv}
\bibliography{references_v1}
\appendix 
%\section{}
\section{Tuning parameters and proposal densities for the PMwG sampler\label{tuningPMwG}}
For  PMwG,  it is necessary to specify the number of particles $R$, and the proposal densities
$m_{j}\left(\boldsymbol{\alpha}_{j}|\boldsymbol{\theta},\boldsymbol{y}_{j}\right)$,
for each subject $j=1,...,S$. \cite{Gunawan2017} use the prior densities $p\left(\boldsymbol{\alpha}_{j}|\boldsymbol{\mu},\boldsymbol{\Sigma}\right)$ 
as the proposal densities for the random effects. The practical performance of the algorithm is
greatly enhanced by choosing efficient proposal densities. To simplify this choice, we develop efficient proposal densities in three stages: burnin, initial adaptation, and sampling. In the burnin and the initial adaptation stages, the proposal density for subject $j$ is a mixture over the prior group-level distribution for the 
random effects, $p\left(\boldsymbol{\alpha}_{j}|\boldsymbol\theta\right)$, and a normal distribution centred on the previous sample for the random effect $N\big(\boldsymbol{\alpha}_{j};\boldsymbol{\alpha}_{j}^{\left(iter-1\right)},
\boldsymbol{\Sigma}\big)$, 
\begin{equation}
m_{j}\left(\boldsymbol{\alpha}_{j}|\boldsymbol{\theta},\boldsymbol{y}_{j}\right)=w_{mix}N\big(\boldsymbol{\alpha}_{j};\boldsymbol{\alpha}_{j}^{\left(iter-1\right)},
\boldsymbol{\Sigma}\big)+\left(1-w_{mix}\right)p\left(\boldsymbol{\alpha}_{j}|\boldsymbol\theta\right),\label{eq:proposal density initial}
\end{equation}
where $\boldsymbol{\alpha}_{j}^{\left(iter-1\right)}$ is the previous iterate
$\boldsymbol{\alpha}_{j}^{k_j}$
for the individual $j$th random effect. In practice, we should use a larger number of particles in the burnin and initial
adaptation stages than in the sampling stages. 

%\footnote{DG: please check this and correct}

%In the burnin and initial adaptation stages we use a random walk as one component of the mixture. 
This proposal can be made more flexible by using  $\epsilon \bs \Sigma $ instead of $\bs \Sigma$, where $0< \epsilon<1$ is a scale factor. In this paper, we set $\epsilon=1$. It is necessary to reduce the scale factor $\epsilon$ when we have larger number of random effects. 
\citet{Gunawan:2019b} estimate a high-dimensional hierarchical LBA model with $S=110$ subjects, $N=1,350$ trials, 
and $D_{\alpha}=30$ individual random effects parameters and set $\epsilon=0.1$. We find empirically in our examples that this strategy works well in practice.
%This strategy works well in practice

In the sampling stage, we use the posterior MCMC draws $\left(\boldsymbol{\alpha}_{1:S},\boldsymbol{\theta}\right)$
from the initial adaptation stage to adaptively build more efficient proposal densities
$m_{j}\left(\boldsymbol{\alpha}_{j}|\boldsymbol{\theta},\boldsymbol{y}_{j}\right)$,
for each subject $j=1,...,S$. This usually allows the use of a much smaller number of particles. We first transform the posterior draws of the
parameters $\boldsymbol{\Sigma}$ so that they all lie on the real
line. The covariance matrix $\boldsymbol{\Sigma}$ is reparameterised
in terms of its Cholesky factorisation $\boldsymbol{\Sigma}=\boldsymbol{L}\boldsymbol{L}^{T}$,
where $\boldsymbol{L}$ is a lower triangular matrix. We also apply a
log transformation for the diagonal elements of $\boldsymbol{L}$, while 
the subdiagonal elements of $\boldsymbol{L}$ are unrestricted.
For each subject, we fit a normal distribution to the vectors formed by joining the posterior draws of $\boldsymbol{\alpha}_{j}$ with $\left(\boldsymbol{\mu},\boldsymbol{L}\right)$ and obtain the
conditional distribution $g\left(\boldsymbol{\alpha}_{j}|\boldsymbol{\mu},\boldsymbol{L}\right)\sim N\left(\boldsymbol{\alpha}_{j};\boldsymbol{\mu}_{j,prop},\boldsymbol{\Sigma}_{j,prop}\right)$
for $j=1,...,S$. The efficient proposal density for subject $j$ is then the two component mixture
\begin{equation}
m_{j}\left(\boldsymbol{\alpha}_{j}|\boldsymbol{\theta},\boldsymbol{y}_{j}\right)=w_{mix}N\left(\boldsymbol{\alpha}_{j};\bs\mu_{j,prop},\bs\Sigma_{j,prop}\right)+\left(1-w_{mix}\right)p\left(\boldsymbol{\alpha}_{j}|\boldsymbol{\theta}\right).\label{eq:proposal density}
\end{equation}
In this paper, the number of particles in the PMwG method is set to $R_{burnin}=R_{adapt}=R_{sampling}=100$; as discussed above, it is sometimes helpful to use more particles in the burnin and initial adaptation stages. We also set $w_{mix}=0.9$. 

We used $500$ burnin iterations, $500$ adaptation iterations, and $10000$ iterations for sampling.
Following \cite{Hesterberg1995}, including the prior density $p\left(\boldsymbol{\alpha}_{j}|\boldsymbol{\theta}\right)$
in \eqref{eq:proposal density initial} and \eqref{eq:proposal density} ensures that the importance weights
are bounded because it is straightforward to show that the density $p\left(\boldsymbol{y}_{j}|\boldsymbol{\theta},\boldsymbol{\alpha}_{j}\right)$ 
is bounded. This 
ensures that the sampler is ergodic; see the online supplement at \url{osf.io/5b4w3} for further details.

%\section*{Appendix}
%\subsection*{Additional results for the data analysis}
%This section reports some additional results for the data analysis in Section \ref{sec:Simulation-Study-and real applications}.

\section{Tuning Parameters and Proposal Densities for \Dt_SMC{} \label{tuningAISIL}}
\Dt_SMC{}  has three tuning parameters: the number of
particles $R$, the number of Markov move steps $L$, and the number
of SMC samples $M$. The bigger the number of SMC samples $M$, the better the approximation to
the posterior density $p\left(\boldsymbol{\theta},\boldsymbol{\alpha}|\boldsymbol{y}\right)$.
By using the result of \citet{DelMoral:2006}, \dt_SMC{}
provides consistent inference for the posterior density $p\left(\boldsymbol{\theta},\boldsymbol{\alpha}|\boldsymbol{y}\right)$
as the number of annealed samples $M$ goes to infinity, for any given
number of particles $R$. The $L$ Markov moves in step (2e) in Algorithm \ref{alg:Generic-AISIL-Algorithm} help to diversify the collection of parameters
and random effects after the resampling in step (2d) so that they better
approximate the tempered target density. In addition, we can obtain more accurate marginal likelihood estimates with larger number of SMC samples $M$, the number of particles $R$, and the number of Markov move steps $L$.
%The AISIL-RE method implements Markov moves based on PMwG sampler with the conditional MC algorithm given in Algorithm~\ref{alg:conditional Monte-Carlo-Algorithm-1}. 

The prior density $p\left(\boldsymbol{\alpha}_{j}|\boldsymbol{\theta}\right)$
is an efficient proposal for the random effects for each subject when the tempering value $a_p$ 
is small as it dominates the tempered density which is quite flat because
$p\left(\boldsymbol{y}_{j}|\boldsymbol{\theta},\boldsymbol{\alpha}_{j}\right)^{a_{p}} \approx 1$.   
In our application we use the prior as a proposal density when $a_p< 0.1$. It is often also adequate 
to use a smaller number of particles $R$ and smaller number of Markov moves $L$ when $a_p$ is small.
When $a_p$ is larger than 0.1, we first fit a normal
distribution -- in the same manner as for the adaptive proposal densities in the PMwG algorithm -- to the current transformed particle cloud $\big\{\left(\boldsymbol{\mu},\boldsymbol{L}\right)_{1:M}^{\left(p\right)},\boldsymbol{\alpha}_{j,1:M}^{\left(p\right)},W_{1:M}^{\left(p\right)}\big\} $
for $j=1,...,S$. 
This gives the conditional distribution 
$g\left(\boldsymbol{\alpha}_{j}|\boldsymbol{\mu},\boldsymbol{L}\right)\sim N\left(\boldsymbol{\alpha}_{j};\boldsymbol{\mu}_{j,prop},\boldsymbol{\Sigma}_{j,prop}\right)$ at each stage of the SMC process. We then use the two
component mixture given in \eqref{eq:proposal density} as a proposal density
and set the mixture weight to $w_{mix}=0.9$. The number of MC samples and Markov moves were set to $R=100$ and $L=10$ respectively. We set $ESS_{T}=0.8M$; i.e., we target an effective sample size of 80\% of the maximum SMC sample size. 
Unlike the PMwG algorithm, \dt_SMC{} does not require an initial adaptation stage to construct the efficient proposal density. Instead, the
proposal densities of the random effects for each subject are obtained from the current particle cloud at each stage of the SMC process.

\section{Details on the Thermodynamic Integration\label{DetailThermodynamicIntegration}}

The first order quadrature approximation $\left(TI_{1}\right)$ to
the integral in \eqref{eq:thermodynamicIdentity} of the main text is based
on the trapezoidal rule and is 
\begin{multline}
\widehat{\log p\left(y\right)}=\sum_{p=1}^{P}\left(\frac{a_{p}-a_{p-1}}{2}\right)\\
\left(\E_{\xi_{a_{p}}}\left(\log\left\{ p\left(\boldsymbol{y}|\boldsymbol{\theta},\boldsymbol{\alpha}\right)\right\} \right)+\E_{\xi_{a_{p-1}}}\left(\log\left\{ p\left(\boldsymbol{y}|\boldsymbol{\theta},\boldsymbol{\alpha}\right)\right\} \right)\right).\label{eq:firstorderTI}
\end{multline}
Discretising the tempering sequence $a_{p}$ and using the trapezoidal
rule introduces some bias to the estimate in  \eqref{eq:firstorderTI}.
\citet{Friel:2014} propose using the corrected trapezium rule method  of
\citet{Atkinson2004} to reduce the bias of $TI_{1}$. The corrected trapezium rule approximates the integral of a function $f$ between points
$a$ and $b$ as 
\begin{equation}
\int_{a}^{b}f\left(x\right)dx\approx \left(b-a\right)\left[\frac{f\left(a\right)+f\left(b\right)}{2}\right]-\frac{\left(b-a\right)^{3}}{12}f^{''}\left(c\right),\label{eq:corrected trapeizodal rule}
\end{equation}
where $c \in \left[a,b\right]$ and $f^{'}\left(\cdot\right) $ and $f^{''}\left(\cdot\right)$
are the first and second derivative of  $f$. The first term in 
\eqref{eq:corrected trapeizodal rule} is the usual trapezium rule approximation 
used in $TI_{1}$. The second derivative $f^{''}\left(c\right)$ in the second term of \eqref{eq:corrected trapeizodal rule}
can be approximated as
\[
f^{''}\left(c\right)\approx\frac{f^{'}\left(b\right)-f^{'}\left(a\right)}{b-a}.
\]
Hence,  
\begin{equation}
\int_{a}^{b}f\left(x\right)dx \approx\left(b-a\right)\left[\frac{f\left(a\right)+f\left(b\right)}{2}\right]-\frac{\left(b-a\right)^{2}}{12}\left(f^{'}\left(b\right)-f^{'}\left(a\right)\right).\label{eq:corrected trapeizodal rule-1}
\end{equation}
Differentiating $E_{\xi_{a_{p}}}\left(\log\left\{ p\left(\boldsymbol{y}|\boldsymbol{\theta},\boldsymbol{\alpha}\right)\right\} \right)$
with respect to $a_{p}$ yields 
\begin{eqnarray}
\frac{d}{dt}E_{\xi_{a_{p}}}\left(\log\left\{ p\left(\boldsymbol{y}|\boldsymbol{\theta},\boldsymbol{\alpha}\right)\right\} \right) & = & \E_{\xi_{a_{p}}}\left(\log\left\{ p\left(\boldsymbol{y}|\boldsymbol{\theta},\boldsymbol{\alpha}\right)\right\} ^{2}\right)-\left(\E_{\xi_{a_{p}}}\left(\log\left\{ p\left(\boldsymbol{y}|\boldsymbol{\theta},\boldsymbol{\alpha}\right)\right\} \right)\right)^{2}\nonumber \\
 & = & \V_{\xi_{a_{p}}}\left(\log\left\{ p\left(\boldsymbol{y}|\boldsymbol{\theta},\boldsymbol{\alpha}\right)\right\} \right).\label{eq:derivativeExpectation}
\end{eqnarray}
Using the results in \eqref{eq:derivativeExpectation}, the
second order quadrature approximation $\left(TI_{2}\right)$ to the integral in \eqref{eq:thermodynamicIdentity} of the main text is 
\begin{multline}
\widehat{\log p\left(y\right)}=\sum_{p=1}^{P}\left(\frac{a_{p}-a_{p-1}}{2}\right)\\
\times\left[\E_{\xi_{a_{p}}}\left(\log\left\{ p\left(\boldsymbol{y}|\boldsymbol{\theta},\boldsymbol{\alpha}\right)\right\} \right)+\E_{\xi_{a_{p-1}}}\left(\log\left\{ p\left(\boldsymbol{y}|\boldsymbol{\theta},\boldsymbol{\alpha}\right)\right\} \right)\right]\\
-\sum_{p=1}^{P}\frac{\left(a_{p}-a_{p-1}\right)^{2}}{12}\\
\left[\V_{\xi_{a_{p}}}\left(\log\left\{ p\left(\boldsymbol{y}|\boldsymbol{\theta},\boldsymbol{\alpha}\right)\right\} \right)-\V_{\xi_{a_{p-1}}}\left(\log\left\{ p\left(\boldsymbol{y}|\boldsymbol{\theta},\boldsymbol{\alpha}\right)\right\} \right)\right],
\end{multline}
where both the expectation $\E_{\xi_{a_{p}}}\left(\log\left\{ p\left(\boldsymbol{y}|\boldsymbol{\theta},\boldsymbol{\alpha}\right)\right\} \right)$
and variance \\ $\V_{\xi_{a_{p}}}\left(\log\left\{ p\left(\boldsymbol{y}|\boldsymbol{\theta},\boldsymbol{\alpha}\right)\right\} \right)$
can be estimated using the \dt_SMC{}  output at the tempering  value $a_{p}$. 
See \ref{DetailThermodynamicIntegration} for more details. 

Estimating the marginal likelihood using thermodynamic integration requires careful consideration of three tuning issues: (i)~the algorithm that samples from the tempered
posterior $\xi_{a_{p}}\left(\boldsymbol{\theta},\boldsymbol{\alpha}|\boldsymbol{y}\right)$,
for $p=0,...,P$; (ii)~the number of tempering steps $\left(P\right)$;
and (iii)~the tempering sequence $a_{p}$ for $p=0,...,P$. \citet{friel2008marginal}
use standard MCMC algorithms, such as Gibbs and Metropolis-Hastings, because they deal with tractable likelihoods; 
\citet{Evans:2019} use DE-MCMC to sample
from the tempered target posterior. It is important to make sure that the MCMC sampler used converges for each value of the tempering sequence
$a_{p}$, $p=0,...,P$ to ensure the accuracy of the marginal likelihood
estimates. However,  it is difficult in general to assess whether the chains
mix adequately and converge to the invariant tempered target density. Furthermore,
standard random walk Metropolis-Hastings and DE-MCMC algorithms usually
suffer from high autocorrelations between samples,
and slow or uncertain convergence for models with a large number of
parameters. The MCMC convergence problems can sometimes be solved
by increasing the number of MCMC samples, but this leads to increased
computational workload and the MCMC algorithms suffer from limited parallelizability. \citet{friel2008marginal}, \citet{Xie2010}, and \citet{Evans:2019}
use the tempering sequence $a_{p}=\left((p-1)/(P-1)\right)^{1/0.3}$,
for $p=0,...,P$. These tempering sequences place more computational effort on temperatures near $0$, where the tempered posterior changes rapidly. They also fix the tempering steps before running the algorithm by experimenting with different numbers of tempering steps and then choose the tempering step $P$ that gives the lowest standard error of the log of marginal likelihood estimates. In contrast, 
our approach through \dt_SMC{} gives a principled, but perhaps not optimal, 
sequence of tempering steps.
%If the number of tempering steps is too small, then the estimate of the marginal likelihood
%will be inaccurate, but if the number tempering steps is too large, this
%also leads to increased computational workload.

\section{Assumptions for the Proposal Densities\label{AssumptionProp}}

We define the support of the posterior and the proposal densities as
\begin{eqnarray*}
\S_{j}^{\boldsymbol{\theta}}\coloneqq\left(\boldsymbol{\alpha}_{j}\in\boldsymbol{\chi}_{\alpha}:p\left(\boldsymbol{\alpha}_{j}|\boldsymbol{\theta},\boldsymbol{y}_{j}\right)>0\right) & \textrm{and} & \Q_{j}^{\boldsymbol{\theta}}\coloneqq\left\{ \boldsymbol{\alpha}_{j}\in\boldsymbol{\chi}_{\alpha}:m_{j}\left(\boldsymbol{\alpha}_{j}|\boldsymbol{\theta},\boldsymbol{y}_{j}\right)>0\right\} .
\end{eqnarray*}
and  assume that $\S_{j}^{\boldsymbol{\theta}}\subseteq \Q_{j}^{\boldsymbol{\theta}}$
for any $\boldsymbol{\theta}\in R^{d_{\theta}} $ and $j=1,...,S$.
This ensures that the $m_{j}\left(\boldsymbol{\alpha}_{j}|\boldsymbol{\theta},\boldsymbol{y}_{j}\right)$
can be used as proposal densities to approximate $p\left(\boldsymbol{\alpha}_{j}|\boldsymbol{\theta},\boldsymbol{y}_{j}\right)$.

\section{Further details on the joint density of response times and response choice\label{DetailjointdensityRTandRE}}
We derive in detail the joint density in \eqref{eq: joint density RE RC} of the response choice $RE$ and response time $RT$.
Let $\mu_1(\cdot)$  be the counting measure on the set $\{2,...,C\}$, i.e. $\mu_1(c)=1$ for $c=2, \dots, C$, and $\mu_2(\cdot)$ be the usual Lebesgue measure on the line.
The joint probability for $RE=c$ and $RT\in(t,t+dt)$ is
\begin{align*}
P\big(RE=c, RT\in(t,t+dt)\big)&=P\big(RE=c\big)P\big(RT\in(t,t+dt)|RE=c\big)\\
&=P(T_k>t,k\not=c)P\big(T_c\in(t,t+dt)\big)\\
&=\prod_{k\not= c}(1-F_k(t))f_c(t)\mu_1(c)\mu_2(dt).
\end{align*}
This implies that \eqref{eq: joint density RE RC} is the joint density of $RE$ and $RT$ with respect to the product measure $\mu_1\otimes\mu_2$.

%We will use  \eqref{eq: joint density RE RC} as the
%definition of the joint density of $RE$ and $RT$ in the expressions  for the posterior density and the
%marginal likelihood, as well as the proposed sampling schemes.
%We note, however,  that to be rigorous it is necessary to define any density function with respect to some
%measure. In this case, we have a bivariate density with respect to the variable $RE$ that is discrete 
%and takes the values $1, \dots, C$ and a second variable that takes 
%continuous values on the positive real line.
%Thus $RE$ is associated with the counting measure $\mu(\cdot)$ such that $\mu(c)=1$ for $c=1, \dots, C$, while $t$ is associated with the usual (Lebesgue) measure on the line which we denote as $d t$. Thus,
%$(RE,RT)$ has the density \eqref{eq: joint density RE RC} with respect %to the product measure $\mu(c) dt$.

\section{Comparing PMwG and DE-MCMC for the simplified LBA model \label{DetailComparisonPMwG}}

This section compares the performance of
PMwG  and  DE-MCMC  applied to a simplified hierarchical LBA model 
that assumes independent normal distributions for the log-transformed individual random effects using the same simulated dataset with $S=100$ and $N=1000$ trials given in Section \ref{subsec:Simulation-Study}. 
%\footnote{I assume that you are taking logs of the random effects here so we should say that} 
This is a standard assumption for DE-MCMC applications of LBA, and simplifies model estimation considerably 
because only the diagonal elements of $\bs \Sigma$ are estimated. 
For each random effect component in $\bs {\alpha}_{j}$,
we define the independent univariate normal distributions 
$\alpha_{dj}  \sim  N\big(\mu_{d},\sigma^2_{d}\big)$, $d=1,...,D_{\alpha}$. We use the same priors for the group level parameters defined in Section \ref{sec:The-Linear-Ballistic accumulator}.
%The prior for $\mu$ was a standard multivariate normal and the prior for the standard deviation was half cauchy. 

%We take the following priors $p\big(\mu_{\alpha_{b^{\left(z\right)}}}\big)\sim N\left(0,3^{2}\right)$
%for $z=1,...,3$, $p\big(\mu_{\alpha_{A}}\big)\sim N\left(1,3^{2}\right)$,
%$p\big(\mu_{\alpha_{v^{\left(1\right)}}}\big)=p\big(\mu_{\alpha_{v^{\left(2\right)}}}\big)\sim N(1,3^{2})$,
%$p\left(\mu_{\tau}\right)\sim N\left(-2,1\right)$, and we take $N\left(0,3^{2}\right)$
%for the log-transformation of the standard deviation parameters.

For DE-MCMC we used $15$ chains and ran each for $5,000$ iterations, 
discarding the first $2,500$ iterations from each chain,
and thinning by keeping only every $10$th draws;  
we obtain a total of $3,750$ samples. To match with DE-MCMC sampler, we generated $4,000$ draws using the PMwG sampler for the analysis of the posterior distribution. The posterior distributions from the two samplers agreed closely, which is to be expected with the large (100 participants, each with $1,000$ trials) and clean (synthetic data generated without noise) sample.

\begin{figure}[h]
\centering{}\includegraphics[width=13cm,height=7cm]{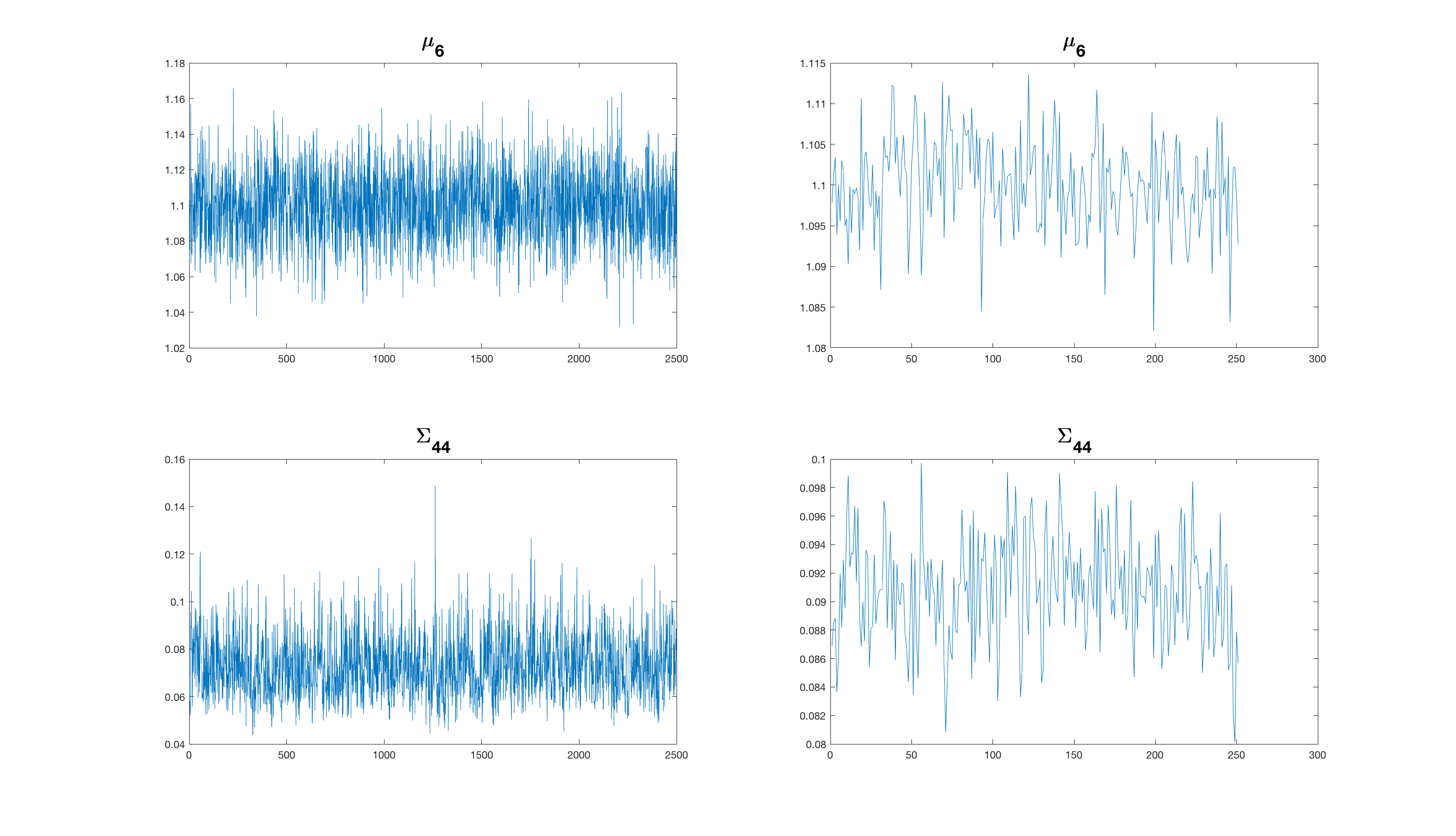}
\caption{Trace plots of the iterates of two of the LBA parameters
estimated; using PMwG (left panels) and thinned (one in ten) DE-MCMC iterates (right panels) for the simplified LBA model, with uncorrelated priors for the random effects. \label{fig:Left-Figures:-Trace plot PG-DEMCMC param}}
\end{figure}

Figure \ref{fig:Left-Figures:-Trace plot PG-DEMCMC param} shows the trace plots of the iterates of 
two  group-level LBA parameters estimated using PMwG and DE-MCMC, respectively. It is clear that even with substantial thinning, the DE-MCMC samples do not mix as well as the (un-thinned) PMwG samples. Similar plots were obtained for the other group level parameters. Since the DE-MCMC is based on multiple interacting chains, the DE-MCMC samples is plotted by taking the mean across iterates from each chain. The impression of poorer mixing for the DE-MCMC sampler was confirmed by calculating IACT values for all parameters. Those calculations showed higher inefficiency factors for the 10x thinned DE-MCMC samples than for the un-thinned PMwG samples (median IACT for parameters was 2.10, compared with 1.19 for the PMwG samples). We also calculated IACT on the chain formed by taking the mean across iterates from each chain. This likely provides an underestimate of inefficiency experienced in practice, where the function calculated from samples is often not summed over chains.   %\footnote{Figure F6 is nice. 
%\begin{itemize}
%    \item 
%In the caption I suggest writing 
%\lq Trace plots of the iterates of two of the LBA parameters
%estimated;  using PMwG (left panels); using thinned (1 in ten)  DE-MCMC iterates (right panels)   for the simplified LBA model \item 
%why not also add another two panels showing the unthinned DE-MCMC iterates ? 
%\item 
%Note I added \lq Similar plots were obtained for the other group level parameters.\rq{} I am assuming it is true.  
%\end{itemize}
%}
%Similar conclusions hold from the trace plot of random effects as well as 
%for the trace plot of the other LBA group-level parameters. 
%\footnote{In figure F.6, please replace by $\mu_1, $ etc as appropriate. }
%The wall-clock computation time for the DE-MCMC sampler was around 180 minutes on a 4-core CPU, with an R implementation.

\section{Estimating Marginal Likelihood using the PMwG output \label{MarginalLikelihoodPMwG}}
\citet{Gronau:2019} and \citet{Evans:2019} use the posterior samples obtained from the DE-MCMC sampler to estimate the marginal likelihood by bridge sampling and thermodynamic integration, respectively. Section \ref{sec:Simulation-Study-and real applications} and \ref{DetailComparisonPMwG} show that the PMwG sampler is more reliable and efficient than the DE-MCMC sampler. Therefore, it is instructive to use the PMwG output to estimate the marginal likelihood by bridge sampling and thermodynamic integration. 

For thermodynamic integration, we first run the PMwG sampler for each value of the tempering sequence $a_{p}$. The posterior samples at each tempering sequence $a_{p}$ is the estimate of $E_{p_{a_p}(\boldsymbol{\theta},\boldsymbol{\alpha}|\boldsymbol{y})}\left(\log\left\{ p\left(\boldsymbol{y}|\boldsymbol{\theta},\boldsymbol{\alpha}\right)\right\} \right)$ and these samples can be used to obtain the marginal likelihood using thermodynamic integration methods described in \ref{DetailThermodynamicIntegration}. 

We can similarly first obtain the posterior samples using the PMwG; and then use these posterior samples to obtain the marginal likelihood estimate by the bridge sampling method as described in \citet{Gronau:2019}.

\end{document}